\documentclass{svmult}
\usepackage{epsfig,graphicx} 
\usepackage[bottom]{footmisc}
\usepackage{natbib,url}      
\bibpunct{(}{)}{;}{a}{}{,}   
\def\cite#1{\citealp{#1}}    
\def\authorindex#1{}  
\def\figspath{.}

\begin{document}\newcount\preprintheader\preprintheader=1


\def\thisvolume{these proceedings}

\def\aj{{AJ}}			
\def\araa{{ARA\&A}}		
\def\apj{{ApJ}}			
\def\apjl{{ApJ}}		
\def\apjs{{ApJS}}		
\def\ao{{Appl.\ Optics}} 
\def\apss{{Ap\&SS}}		
\def\aap{{A\&A}}		
\def\aapr{{A\&A~Rev.}}		
\def\aaps{{A\&AS}}		
\def\an{{Astron.\ Nachrichten}}
\def\aspcs{{ASP Conf.\ Ser.}}
\def\assp{{Astrophys.\ \& Space Sci.\ Procs., Springer, Heidelberg}}
\def\azh{{AZh}}			
\def\baas{{BAAS}}		
\def\jrasc{{JRASC}}	
\def\memras{{MmRAS}}		
\def\mnras{{MNRAS}}
\def\nat{{Nat}}		
\def\pra{{Phys.\ Rev.\ A}} 
\def\prb{{Phys.\ Rev.\ B}}		
\def\prc{{Phys.\ Rev.\ C}}		
\def\prd{{Phys.\ Rev.\ D}}		
\def\prl{{Phys.\ Rev.\ Lett.}} 
\def\pasp{{PASP}}
\def\pasj{{PASJ}}		
\def\qjras{{QJRAS}}
\def\science{{Sci}}		
\def\skytel{{S\&T}}		
\def\solphys{{Solar\ Phys.}} 
\def\sovast{{Soviet\ Ast.}}  
\def\ssr{{Space\ Sci.\ Rev.}}
\def\svassp{{Astrophys.\ Space Sci.\ Procs., Springer, Heidelberg}}
\def\zap{{ZAp}}			
\let\astap=\aap
\let\apjlett=\apjl
\let\apjsupp=\apjs
\def\grl{{Geophys.\ Res.\ Lett.}}  
\def\jgr{{J. Geophys.\ Res.}} 

\def\ion#1#2{{\rm #1}\,{\uppercase{#2}}}  
\def\deg{\hbox{$^\circ$}}
\def\sun{\hbox{$\odot$}}
\def\earth{\hbox{$\oplus$}}
\def\la{\mathrel{\hbox{\rlap{\hbox{\lower4pt\hbox{$\sim$}}}\hbox{$<$}}}}
\def\ga{\mathrel{\hbox{\rlap{\hbox{\lower4pt\hbox{$\sim$}}}\hbox{$>$}}}}
\def\sq{\hbox{\rlap{$\sqcap$}$\sqcup$}}
\def\arcmin{\hbox{$^\prime$}}
\def\arcsec{\hbox{$^{\prime\prime}$}}
\def\fd{\hbox{$.\!\!^{\rm d}$}}
\def\fh{\hbox{$.\!\!^{\rm h}$}}
\def\fm{\hbox{$.\!\!^{\rm m}$}}
\def\fs{\hbox{$.\!\!^{\rm s}$}}
\def\fdg{\hbox{$.\!\!^\circ$}}
\def\farcm{\hbox{$.\mkern-4mu^\prime$}}
\def\farcs{\hbox{$.\!\!^{\prime\prime}$}}
\def\fp{\hbox{$.\!\!^{\scriptscriptstyle\rm p}$}}
\def\micron{\hbox{$\mu$m}}
\def\onehalf{\hbox{$\,^1\!/_2$}}	
\def\onethird{\hbox{$\,^1\!/_3$}}
\def\twothirds{\hbox{$\,^2\!/_3$}}
\def\onequarter{\hbox{$\,^1\!/_4$}}
\def\threequarters{\hbox{$\,^3\!/_4$}}
\def\ubv{\hbox{$U\!BV$}}		
\def\ubvr{\hbox{$U\!BV\!R$}}		
\def\ubvri{\hbox{$U\!BV\!RI$}}		
\def\ubvrij{\hbox{$U\!BV\!RI\!J$}}		
\def\ubvrijh{\hbox{$U\!BV\!RI\!J\!H$}}		
\def\ubvrijhk{\hbox{$U\!BV\!RI\!J\!H\!K$}}		
\def\ub{\hbox{$U\!-\!B$}}		
\def\bv{\hbox{$B\!-\!V$}}		
\def\vr{\hbox{$V\!-\!R$}}		
\def\ur{\hbox{$U\!-\!R$}}


\def\labelitemi{{\bf --}}  

\def\rmit#1{{\it #1}}              
\def\rmit#1{{\rm #1}}              
\def\etal{\rmit{et al.}}           
\def\etc{\rmit{etc.}}           
\def\ie{\rmit{i.e.,}}              
\def\eg{\rmit{e.g.,}}              
\def\cf{cf.}                       
\def\viz{\rmit{viz.}}
\def\vs{\rmit{vs.}}

\def\rot{\hbox{\rm rot}}
\def\div{\hbox{\rm div}}
\def\lesssim{\mathrel{\hbox{\rlap{\hbox{\lower4pt\hbox{$\sim$}}}\hbox{$<$}}}}
\def\gtrsim{\mathrel{\hbox{\rlap{\hbox{\lower4pt\hbox{$\sim$}}}\hbox{$>$}}}}
\def\mathstacksym#1#2#3#4#5{\def#1{\mathrel{\hbox to 0pt{\lower 
    #5\hbox{#3}\hss} \raise #4\hbox{#2}}}}
\mathstacksym\lesssim{$<$}{$\sim$}{1.5pt}{3.5pt} 
\mathstacksym\gtrsim{$>$}{$\sim$}{1.5pt}{3.5pt} 
\mathstacksym\lrarrow{$\leftarrow$}{$\rightarrow$}{2pt}{1pt} 
\mathstacksym\lessgreat{$>$}{$<$}{3pt}{3pt} 

\def\dif{\: {\rm d}}                       
\def\ep{\:{\rm e}^}                        
\def\dash{\hbox{$\,-\,$}}                  
\def\is{\!=\!}                             

\def\starname#1#2{${#1}$\,{\rm {#2}}}  
\def\Teff{\hbox{$T_{\rm eff}$}}   

\def\kms{\hbox{km$\;$s$^{-1}$}}
\def\ms{\hbox{m$\;$s$^{-1}$}}
\def\Mxcm{\hbox{Mx\,cm$^{-2}$}}    

\def\Bapp{\hbox{$B_{\rm app}$}}    

\def\komega{($k, \omega$)}                 
\def\kf{($k_h,f$)}                         
\def\VminI{\hbox{$V\!\!-\!\!I$}}           
\def\IminI{\hbox{$I\!\!-\!\!I$}}           
\def\VminV{\hbox{$V\!\!-\!\!V$}}           
\def\Xt{\hbox{$X\!\!-\!t$}}                

\def\level #1 #2#3#4{$#1 \: ^{#2} \mbox{#3} ^{#4}$}   

\def\specchar#1{\uppercase{#1}}    
\def\AlI{\mbox{Al\,\specchar{i}}}  
\def\BI{\mbox{B\,\specchar{i}}} 
\def\BII{\mbox{B\,\specchar{ii}}}  
\def\BaI{\mbox{Ba\,\specchar{i}}}  
\def\BaII{\mbox{Ba\,\specchar{ii}}} 
\def\CI{\mbox{C\,\specchar{i}}} 
\def\CII{\mbox{C\,\specchar{ii}}} 
\def\CIII{\mbox{C\,\specchar{iii}}} 
\def\CIV{\mbox{C\,\specchar{iv}}} 
\def\CaI{\mbox{Ca\,\specchar{i}}} 
\def\CaII{\mbox{Ca\,\specchar{ii}}} 
\def\CaIII{\mbox{Ca\,\specchar{iii}}} 
\def\CoI{\mbox{Co\,\specchar{i}}} 
\def\CrI{\mbox{Cr\,\specchar{i}}} 
\def\CriI{\mbox{Cr\,\specchar{ii}}} 
\def\CsI{\mbox{Cs\,\specchar{i}}} 
\def\CsII{\mbox{Cs\,\specchar{ii}}} 
\def\CuI{\mbox{Cu\,\specchar{i}}} 
\def\FeI{\mbox{Fe\,\specchar{i}}} 
\def\FeII{\mbox{Fe\,\specchar{ii}}} 
\def\FeIX{\mbox{Fe\,\specchar{ix}}}
\def\FeX{\mbox{Fe\,\specchar{x}}}
\def\FeXVI{\mbox{Fe\,\specchar{xvi}}}
\def\FrI{\mbox{Fr\,\specchar{i}}}
\def\HI{\mbox{H\,\specchar{i}}} 
\def\HII{\mbox{H\,\specchar{ii}}} 
\def\Hmin{\hbox{\rmH$^{^{_{\scriptstyle -}}}$}}      
\def\Hemin{\hbox{{\rm He}$^{^{_{\scriptstyle -}}}$}} 
\def\HeI{\mbox{He\,\specchar{i}}} 
\def\HeII{\mbox{He\,\specchar{ii}}} 
\def\HeIII{\mbox{He\,\specchar{iii}}} 
\def\KI{\mbox{K\,\specchar{i}}} 
\def\KII{\mbox{K\,\specchar{ii}}} 
\def\KIII{\mbox{K\,\specchar{iii}}} 
\def\LiI{\mbox{Li\,\specchar{i}}} 
\def\LiII{\mbox{Li\,\specchar{ii}}} 
\def\LiIII{\mbox{Li\,\specchar{iii}}} 
\def\MgI{\mbox{Mg\,\specchar{i}}} 
\def\MgII{\mbox{Mg\,\specchar{ii}}} 
\def\MgIII{\mbox{Mg\,\specchar{iii}}} 
\def\MnI{\mbox{Mn\,\specchar{i}}} 
\def\NI{\mbox{N\,\specchar{i}}}
\def\NIV{\mbox{N\,\specchar{iv}}}
\def\NaI{\mbox{Na\,\specchar{i}}}
\def\NaII{\mbox{Na\,\specchar{ii}}}
\def\NaIII{\mbox{Na\,\specchar{iii}}}
\def\NeVIII{\mbox{Ne\,\specchar{viii}}} 
\def\NiI{\mbox{Ni\,\specchar{i}}} 
\def\NiII{\mbox{Ni\,\specchar{ii}}}
\def\NiIII{\mbox{Ni\,\specchar{iii}}} 
\def\OI{\mbox{O\,\specchar{i}}} 
\def\OVI{\mbox{O\,\specchar{vi}}}
\def\RbI{\mbox{Rb\,\specchar{i}}} 
\def\SII{\mbox{S\,\specchar{ii}}} 
\def\SiI{\mbox{Si\,\specchar{i}}} 
\def\SiII{\mbox{Si\,\specchar{ii}}} 
\def\SrI{\mbox{Sr\,\specchar{i}}}
\def\SrII{\mbox{Sr\,\specchar{ii}}}
\def\TiI{\mbox{Ti\,\specchar{i}}} 
\def\TiII{\mbox{Ti\,\specchar{ii}}} 
\def\TiIII{\mbox{Ti\,\specchar{iii}}} 
\def\TiIV{\mbox{Ti\,\specchar{iv}}} 
\def\VI{\mbox{V\,\specchar{i}}} 
\def\HtwoO{\mbox{H$_2$O}}        
\def\Otwo{\mbox{O$_2$}}          

\def\Halpha{\mbox{H\hspace{0.1ex}$\alpha$}} 
\def\Ha{\mbox{H\hspace{0.2ex}$\alpha$}}
\def\Hbeta{\mbox{H\hspace{0.2ex}$\beta$}}
\def\Hgamma{\mbox{H\hspace{0.2ex}$\gamma$}}
\def\Hdelta{\mbox{H\hspace{0.2ex}$\delta$}}
\def\Hepsilon{\mbox{H\hspace{0.2ex}$\epsilon$}}
\def\Hzeta{\mbox{H\hspace{0.2ex}$\zeta$}}
\def\Lyalpha{\mbox{Ly$\hspace{0.2ex}\alpha$}}
\def\Lybeta{\mbox{Ly$\hspace{0.2ex}\beta$}}
\def\Lygamma{\mbox{Ly$\hspace{0.2ex}\gamma$}}
\def\Lycont{\mbox{Ly\hspace{0.2ex}{\small cont}}}
\def\Baalpha{\mbox{Ba$\hspace{0.2ex}\alpha$}}
\def\Babeta{\mbox{Ba$\hspace{0.2ex}\beta$}}
\def\Bacont{\mbox{Ba\hspace{0.2ex}{\small cont}}}
\def\Paalpha{\mbox{Pa$\hspace{0.2ex}\alpha$}}
\def\Bralpha{\mbox{Br$\hspace{0.2ex}\alpha$}}

\def\NaD{\mbox{Na\,\specchar{i}\,D}}    
\def\NaDone{\mbox{Na\,\specchar{i}\,\,D$_1$}}
\def\NaDtwo{\mbox{Na\,\specchar{i}\,\,D$_2$}}
\def\NaID{\mbox{Na\,\specchar{i}\,\,D}}
\def\NaIDone{\mbox{Na\,\specchar{i}\,\,D$_1$}}
\def\NaIDtwo{\mbox{Na\,\specchar{i}\,\,D$_2$}}
\def\Done{\mbox{D$_1$}}
\def\Dtwo{\mbox{D$_2$}}

\def\Mgbone{\mbox{Mg\,\specchar{i}\,b$_1$}}
\def\Mgbtwo{\mbox{Mg\,\specchar{i}\,b$_2$}}
\def\Mgbthree{\mbox{Mg\,\specchar{i}\,b$_3$}}
\def\MgIb{\mbox{Mg\,\specchar{i}\,b}}
\def\MgIbone{\mbox{Mg\,\specchar{i}\,b$_1$}}
\def\MgIbtwo{\mbox{Mg\,\specchar{i}\,b$_2$}}
\def\MgIbthree{\mbox{Mg\,\specchar{i}\,b$_3$}}

\def\CaIIK{\mbox{Ca\,\specchar{ii}\,K}}       
\def\CaIIH{\mbox{Ca\,\specchar{ii}\,H}}
\def\CaIIHK{\mbox{Ca\,\specchar{ii}\,H\,\&\,K}}
\def\HK{\mbox{H\,\&\,K}}
\def\Kthree{\mbox{K$_3$}}      
\def\Hthree{\mbox{H$_3$}}
\def\Ktwo{\mbox{K$_2$}}
\def\Htwo{\mbox{H$_2$}}
\def\Kone{\mbox{K$_1$}}     
\def\Hone{\mbox{H$_1$}}     
\def\KtwoV{\mbox{K$_{2V}$}}
\def\KtwoR{\mbox{K$_{2R}$}}
\def\KoneV{\mbox{K$_{1V}$}}
\def\KoneR{\mbox{K$_{1R}$}}
\def\HtwoV{\mbox{H$_{2V}$}}
\def\HtwoR{\mbox{H$_{2R}$}}
\def\HoneV{\mbox{H$_{1V}$}}
\def\HoneR{\mbox{H$_{1R}$}}

\def\hk{\mbox{h\,\&\,k}}
\def\kthree{\mbox{k$_3$}}    
\def\hthree{\mbox{h$_3$}}
\def\ktwo{\mbox{k$_2$}}
\def\htwo{\mbox{h$_2$}}
\def\kone{\mbox{k$_1$}}     
\def\hone{\mbox{h$_1$}}     
\def\ktwoV{\mbox{k$_{2V}$}}
\def\ktwoR{\mbox{k$_{2R}$}}
\def\koneV{\mbox{k$_{1V}$}}
\def\koneR{\mbox{k$_{1R}$}}
\def\htwoV{\mbox{h$_{2V}$}}
\def\htwoR{\mbox{h$_{2R}$}}
\def\honeV{\mbox{h$_{1V}$}}
\def\honeR{\mbox{h$_{1R}$}}

\ifnum\preprintheader=1     
\makeatletter  
\def\@maketitle{\newpage
\markboth{}{}%
  {\mbox{} \vspace*{-8ex} \par 
   \em \footnotesize To appear in ``Magnetic Coupling between the Interior 
       and the Atmosphere of the Sun'', eds. S.~S.~Hasan and R.~J.~Rutten, 
       Astrophysics and Space Science Proceedings, Springer-Verlag, 
       Heidelberg, Berlin, 2009.} \vspace*{-5ex} \par
 \def\lastand{\ifnum\value{@inst}=2\relax
                 \unskip{} \andname\
              \else
                 \unskip \lastandname\
              \fi}%
 \def\and{\stepcounter{@auth}\relax
          \ifnum\value{@auth}=\value{@inst}%
             \lastand
          \else
             \unskip,
          \fi}%
  \raggedright
 {\Large \bfseries\boldmath
  \pretolerance=10000
  \let\\=\newline
  \raggedright
  \hyphenpenalty \@M
  \interlinepenalty \@M
  \if@numart
     \chap@hangfrom{}
  \else
     \chap@hangfrom{\thechapter\thechapterend\hskip\betweenumberspace}
  \fi
  \ignorespaces
  \@title \par}\vskip .8cm
\if!\@subtitle!\else {\large \bfseries\boldmath
  \vskip -.65cm
  \pretolerance=10000
  \@subtitle \par}\vskip .8cm\fi
 \setbox0=\vbox{\setcounter{@auth}{1}\def\and{\stepcounter{@auth}}%
 \def\thanks##1{}\@author}%
 \global\value{@inst}=\value{@auth}%
 \global\value{auco}=\value{@auth}%
 \setcounter{@auth}{1}%
{\lineskip .5em
\noindent\ignorespaces
\@author\vskip.35cm}
 {\small\institutename\par}
 \ifdim\pagetotal>157\p@
     \vskip 11\p@
 \else
     \@tempdima=168\p@\advance\@tempdima by-\pagetotal
     \vskip\@tempdima
 \fi
}
\makeatother     
\fi

\title*{\bf Vainu Bappu Memorial Lecture: What is a sunspot?}

\author{D. O. Gough\inst{1,2}}

\authorindex{Gough, D. O.} 

\institute{Institute of Astronomy,  University of Cambridge, UK
    \and
    Department of Applied Mathematics and 
    Theoretical Physics, University of Cambridge, UK}

\maketitle

\setcounter{footnote}{0}

\begin{abstract} 
  Sunspots have been known in the West since Galileo Galilei and
  Thomas Harriot first used telescopes to observe the Sun nearly four
  centuries ago; they have been known to the Chinese for more than two
  thousand years.  They appear as relatively dark patches on the
  surface of the Sun, and are caused by concentrations of magnetism
  which impede the flow of heat from deep inside the Sun up to its
  othewise brilliant surface.  The spots are not permanent: the total
  number of spots on the Sun varies cyclically in time, with a period
  of about eleven years, associated with which there appear to be
  variations in our climate.  When there are many spots, it is more
  dangerous for spacecraft to operate.  The cause of the spots is not
  well understood; nor is it known for sure how they die.  Their
  structure beneath the surface of the Sun is in some dispute,
  although much is known about their properties at the surface,
  including an outward material flow which was discovered by John
  Evershed observing the Sun from Kodaikanal a hundred years ago.  I
  shall give you a glimpse of how we are striving to deepen our
  understanding of these fascinating features, and of some of the
  phenomena that appear to be associated with them.
\end{abstract}

\section{Introduction}     

Sunspots are dark blotches apparent on the surface of the Sun which,
under suitable conditions, such as when the Sun is seen through a
suitably thin cloud, can sometimes be seen with the naked eye.
Reports from China date back more than two thousand years, but in the
West the history is less clear.  It is likely that the pre-Socratic
Greek philosopher Anaxagoras observed sunspots with the naked eye, and
there have been scattered reports of sightings in the literature
since.  In 1607 Johannes Kepler tried to observe with a camera obscura
a transit of Mercury that he had predicted, and did indeed see a dark
spot which he believed to be Mercury, but it is likely that what he
saw was actually a sunspot.

\begin{figure}  
\begin{center}
  \includegraphics[width=0.9\textwidth]{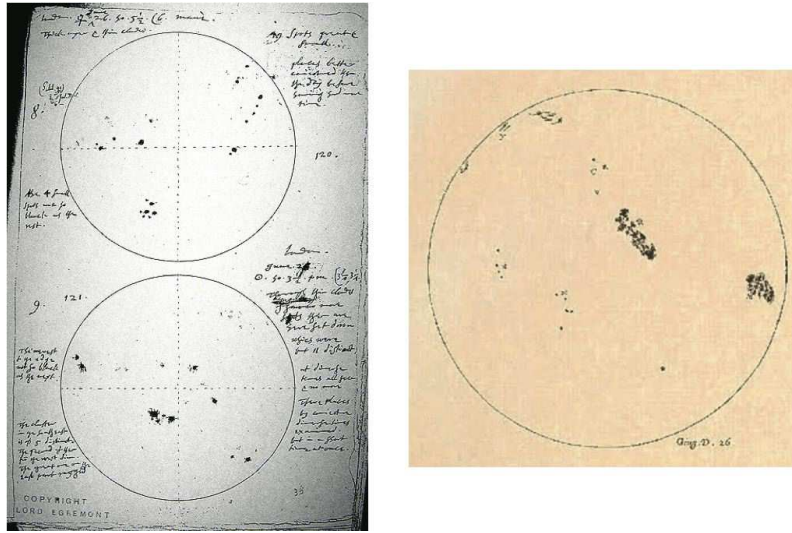}
  \caption[]{On the left is Harriot's sunspot drawing of December
    1610.  On the right is one of a sequence of drawings by Galileo,
    which demonstrates the rotation of the Sun; the rotation is very
    clearly displayed when the drawings are projected in quick
    succession, as in a movie.  It is then evident that the axis of
    rotation is diagonal in the image: from bottom left to top right.
    It is also evident that the sunspots lie in two latitudinal bands
    roughly equidistant from the equator.}
\end{center}
\end{figure}

The scientific study of sunspots began when Thomas Harriot and Galileo
Galilei independently observed the Sun through telescopes late in
1610.  The following year David Fabricius, who had made the first
discovery of a periodic variable star, namely Mira, together with his
son Johannes, also observed spots with a telescope, and published on
them in the autumn of that year.  They had tracked the passage of the
spots across the solar disc, and noticed their reappearance on the
eastern limb a dozen or so days after they had disappeared to the
west, and inferred that the Sun was rotating, a notion that had
already been entertained by Giordano Bruno and Kepler.  Christoph
Scheiner began a serious study at that time: believing the Sun to be
perfect, he attributed the spots to solar satellites which appeared
dark when they passed in front of the disc.  In contrast, with the
help of his prot\'eg\'e Benedetto Castelli, who developed the method
of projecting the Sun's image onto a screen where it could be studied
in great detail, Galileo inferred that the cloud-like spots were
actually on the surface of the Sun, blemishes on what others believed
to be a perfect object, thereby criticizing Scheiner's premise.  The
spots were not permanent features on the surface; nor were their
lifetimes all the same.  A large spot might last a rotation period or
two, after which it disappears, perhaps to be replaced by a spot at a
different location.  Smaller spots are shorter-lived. Galileo also
disagreed with Scheiner's adherence to a geocentric cosmology, having
been rightly convinced by Copernicus's cogent arguments.  The two men,
though civil at first, subsequently became enemies.

Scheiner published a massive book, Rosa Ursina, which became the
standard work on sunspots for a century or more.  By that time he had
at least shed his belief in an unblemished Sun, accepting that the
spots were on the Sun's surface, and by careful measurement of the
motion of the spots he was able to ascertain that the axis of the
Sun's rotation was inclined by about 7$\rm^o$ to the normal to the
plane of the ecliptic.  But he continued to uphold his Ptolemaic
viewpoint.

Further productive work was hampered by a dearth of sunspots
throughout the second half of the seventeenth century, an epoch now
known as the Maunder Minimum.  Perhaps the most important discovery
immediately after that period was by Alexander Wilson in 1769, who
realized from the changing appearance of a spot as it approaches the
solar limb that the central dark umbra is lower than its surroundings,
a phenomenon now known as the Wilson depression.

\section{Subsequent Milestones of Discovery}

An extremely important milestone for the whole of astronomy is Joseph
von Fraunhofer's introduction of spectroscopy, which has enabled
astronomers to draw conclusions about the physical conditions and
chemical composition of celestial objects, most notably the Sun, and
to recognize and measure Doppler wavelength shifts to determine
line-of-sight velocity.  We now know from spectroscopy that sunspots
are cooler than the surrounding photosphere, more of which I shall
discuss later.

\begin{figure}  
\begin{center}
  \includegraphics[width=0.7\textwidth]{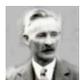}
\caption[]{Landmarks in sunspot discovery} 
\end{center}
\end{figure}

In the few decades after the discovery of sunspots in the West it was
recognized that the number of spots varied with time.  And then there
was the Maunder Minimum -- more than half a century with almost no
spots, an epoch when the appearance of but a single spot was worthy of
comment.  After the reappearance of spots at the beginning of the
eighteenth century, sunspot numbers were again quite variable.  Nobody
at the time appears to have noticed any pattern.  Indeed, it was not
until 1843 that the amateur astronomer Heinrich Schwabe pointed out a
cyclicity, with an estimated period of about 10 years, although
further work revealed that the intervals between successive maxima
vary from 9 to 11.5 years, with an average of about 10.8 years.

\begin{figure}  
\begin{center}
  \includegraphics[width=0.6\textwidth]{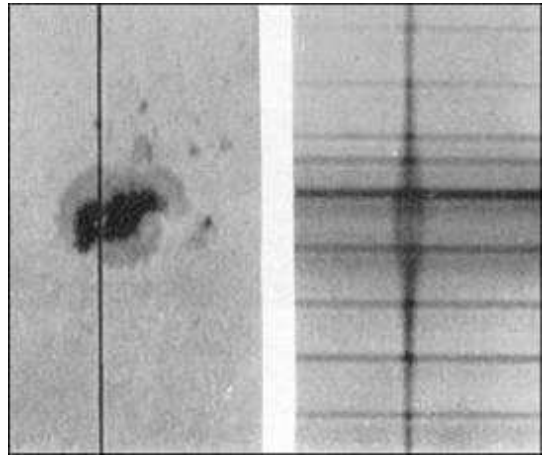}
  \caption[]{ The right-hand panel is a Fraunhofer line in the 
    spectrum of light passed through a slit lying across a sunspot,
    indicated in the left-hand panel, in a portion of the solar image
    not far from disc centre.  The line is split by the magnetic
    field, by an amount which is proportional to the intensity of 
    the field.  Notice
    that the field intensity is roughly uniform in the umbra,
    and then declines gradually to imperceptibility through the
    penumbra.  This is consistent with the sketch reproduced in Fig.~9.}
\end{center}
\end{figure}

In 1908, George Ellery Hale, the man who pioneered astrophysics as a
science beyond the mere identification and plotting of stars, first
observed and recognized Zeeman splitting in sunspots, and so
established the magnetic nature of the spots.  The vertical field is
strongest in the central darkest regions of the spot, where the
strength is about 3000G, and declines gradually outwards (Fig.~3).
Why should such a field concentration come about, and what maintains
it?  Hale subsequently led an investigation into the polarity of
sunspots: large sunspots usually occur in pairs, one leading the other
as the Sun rotates, with the polarity of all leaders being the same in
any hemisphere, but oppositely directed in the northern and southern
hemispheres, and with that polarity changing each sunspot cycle
(producing a magnetic cycle of duration about 22 years).  These
properties are now called Hale's polarity laws.  The presence of a
concentrated magnetic field is now known to be what causes the spot to
exist.  Precisely how the field became so concentrated is less clear.

Some obvious questions come to mind:

$\bullet$ How do sunspots form?\\
\indent $\bullet$ Why are sunspots dark?\\
\indent $\bullet$ What is their structure?\\
\indent $\bullet$ What holds the field together?\\
\indent $\bullet$ How long do sunspots live, and what determines 
  the lifetime?\\
\indent $\bullet$ What is their global effect on the Sun? ... and why?\\
\indent $\bullet$ What causes the sunspot cycle?\\
\indent $\bullet$ Is it predictable?\\

In this lecture I shall address these questions, some of them only
quite cursorily (and not in the order listed), but I shall not be able
to provide satisfactory answers to them all.

\section{Superficial Sunspot Structure}

Figure~4 is a photograph of a sunspot.  There is a central very dark
(in comparison with the normal photosphere) region called the umbra,
which is surrounded by a less dark annulus called the penumbra.
Beyond the penumbra one can see the granulation pattern of convection
in the normal photosphere.  With appropriate exposure, some intensity
variation is visible in the umbra: typically small bright temporally
varying bright dots against a less variable darker background.

\begin{figure}  
\begin{center}
  \includegraphics[width=0.8\textwidth]{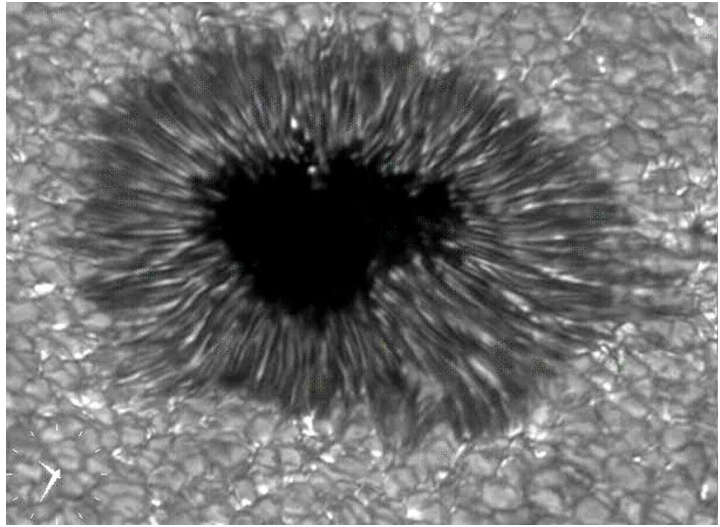}
  \caption[]{Photograph of a sunspot in the G band taken through the
    Dutch Open Telescope.}
\end{center}
\end{figure}

Fine structure in the penumbra is more evident.  It consists mainly of
light and dark filaments radiating from the umbra, apparently aligned
with the magnetic field.  There are also elongated bright regions
aligned with the filaments which extend through only part of the
penumbra; they are called penumbral grains.  Fig.~4 is a single
frame of a movie; when the movie is played it can be seen that the
grains move along the filaments, predominantly inwards in the inner
regions of the penumbra near the umbra, predominantly outwards in the
outer regions.

Doppler observations of weak photospheric spectrum lines reveal a
radially outward flow in the penumbra, the velocity increasing with
radius out to the sunspot boundary.  This is the discovery of John
Evershed, in 1909, to which this conference is dedicated.  In stronger
lines formed in the chromosphere above the photosphere, a reverse flow
is observed.

Sunspots are to be found in a variety of sizes; a medium spot is not
very different in size from the Earth (see Fig.~10).

\section{The Sunspot Cycle}

I have already mentioned that the sunspot number varies cyclically,
with a cycle time of $10.8 \pm 0.9$ years.  Fig.~5 depicts the
variation of a measure of sunspot number (area)\footnote{Rudolf Wolf
  invented a measure of sunspot number which he called `relative
  sunspot number', and which is now called the Wolf, or Z\"urich,
  sunspot number.  It is approximately proportional to an effective
  proportion of the area of the solar disc occupied by sunspots, and,
  since the intensity of sunspot fields does not vary very much from
  one spot to another, it provides an estimate of the total (unsigned)
  magnetic flux emerging from sunspots.} with time since the Maunder
Minimum, with some pre-minimum estimates from the time of Galileo and
Scheiner.  There is proxy evidence that the post-minimum cycle is a
continuation of similar cyclic behaviour occurring before the Maunder
Minimum, with some hint that phase was maintained between them, to the
extent that phase is maintained at all.  Fig.~6 illustrates not only
the variation of sunspot area but also the latitudes at which the
spots occur.  At a typical epoch sunspots are concentrated mainly in
latitudinal belts located roughly symmetrically about the
equator. Spots first appear at the beginning of a cycle in the
vicinity of latitudes $\pm 30\rm{^o}$; as the cycle progresses the
belts migrate equatorwards and eventually merge and disappear as new
belts of reverse magnetic polarity emerge at `high' 
latitudes at the start of the next cycle.  That plot is now called the
butterfly diagram, a name which hardly needs explanation.

\begin{figure}  
\begin{center}
  \includegraphics[width=0.9\textwidth]{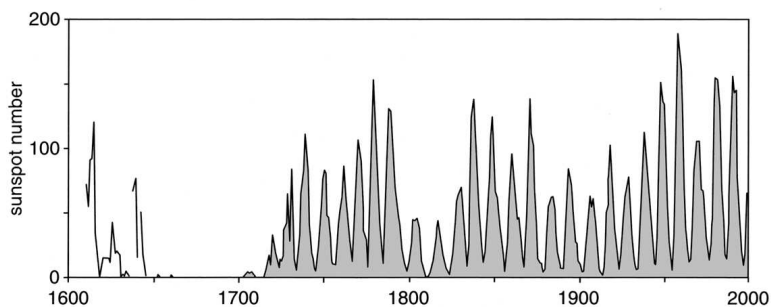}
  \caption[]{Smoothed plot of sunspot numbers through the last three 
    complete centuries.}
\end{center}
\end{figure}

Associated with the sunspot coverage is a variation of solar
irradiance, the total radiative flux from the Sun in the plane of the
ecliptic, normalized to a distance of one astronomical unit.
Irradiance is thus an indicator of the flux of radiation from the Sun
in the direction of the Earth.  The irradiance has been measured
accurately only since detectors could be raised above (most of) the
Earth's atmosphere.  Fig.~7 shows measurements by a variety of
instruments.  It should be appreciated that it is very difficult to
make an absolute measurement, as is evident in the upper panel of the
figure, but if the zero points of the fluxes are shifted appropriately
the measurements can be made to lie on top of each other.  The lower
panel is a (weighted) composite of the shifted curves; the thick line
is a running average, which shows quite clearly a roughly eleven-year
cycle, as one might expect.  Interestingly, comparison with Fig.~6
reveals that at sunspot maximum, when one would expect the greatest
reduction of light output by the dark spots, the irradiance too is a
maximum, as are the magnitudes of the fluctuations in the absolute
sunspot number.  That demands explanation.  One comforting property of
the plot is that at sunspot maximum the fluctuations in the irradiance
are also at their greatest.  We now know that these fluctuations are
caused principally by sunspots and their immediate surroundings moving
into and out of view as the Sun rotates.

\begin{figure}  
\centering
  \includegraphics[width=\textwidth]{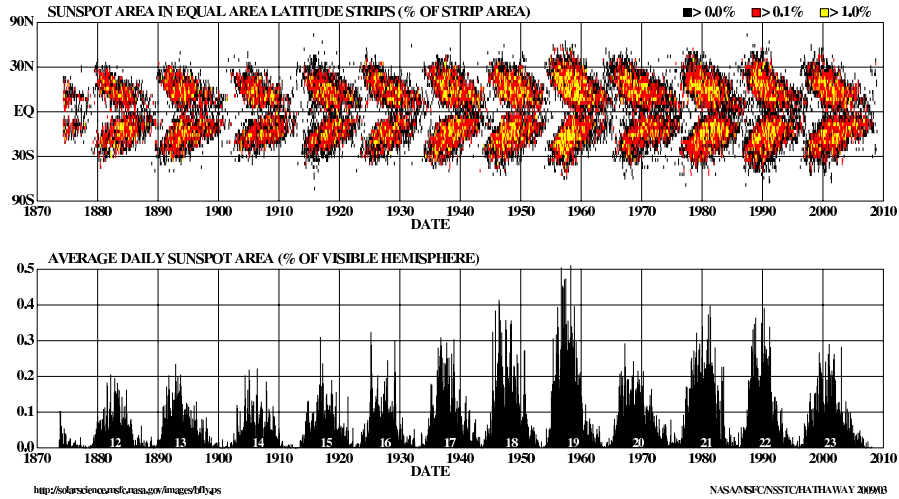}
  \caption[]{The lower panel depicts daily sunspot area, annual
    averages of which correspond to the last 130 years or so of
    Fig.~5.  The upper panel marks the latitudes of the spots
    (compiled by David Hathaway).}
\end{figure}
\begin{figure}  
\begin{center}
  \includegraphics[width=\textwidth]{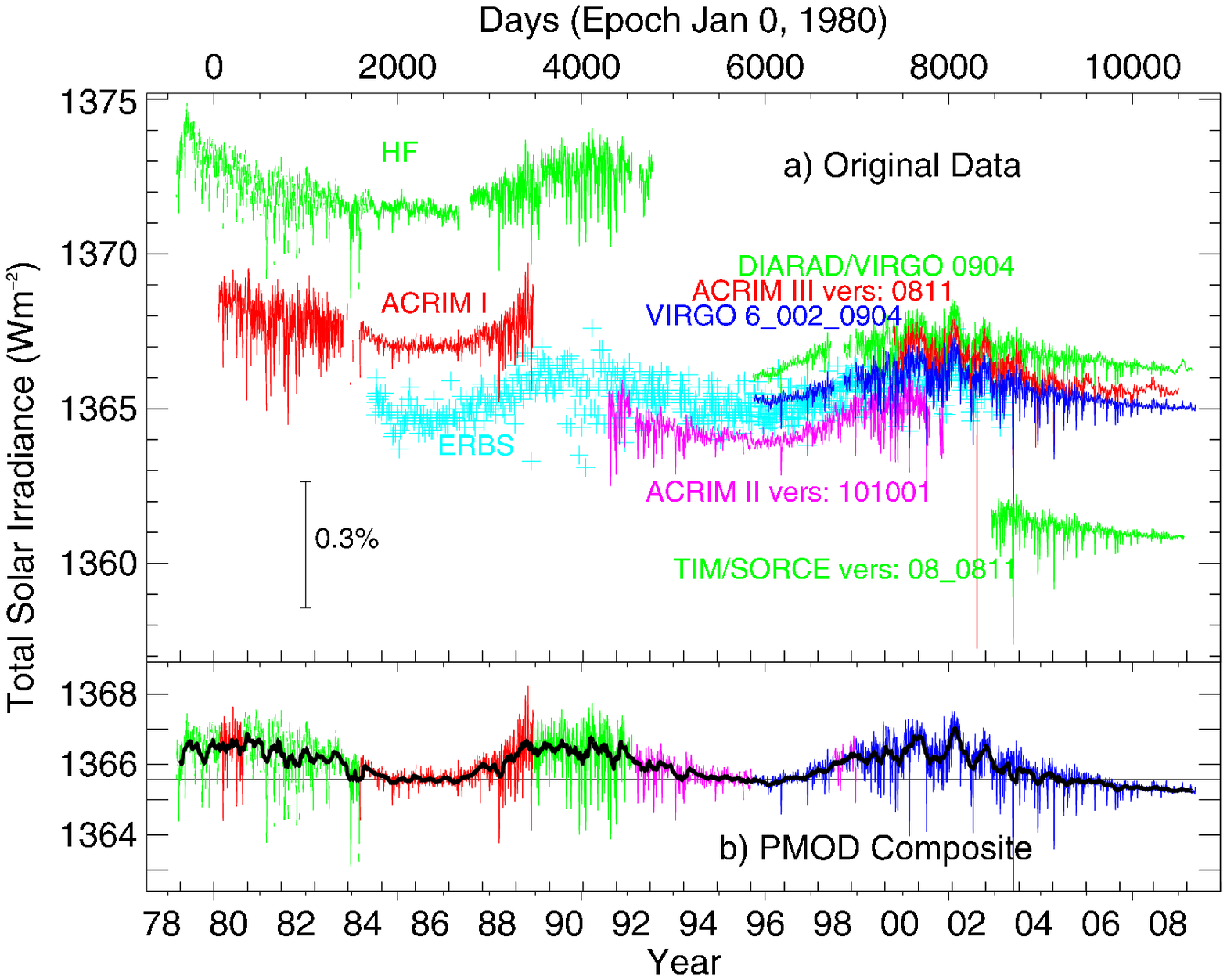}
  \caption[]{Measurements of solar irradiance by several different
    instruments.  In the panel below is a combination of those
    measurements obtained by shifting the zero points to make the
    results lie on top of each other.  The thick superposed line is a
    running mean (Physikalisch-Meteorologisches Observatorium, Davos)}
\end{center}
\end{figure}

Another property evident in Figs.~5 and 6 is that there is a
variation in the value of the sunspot number from one maximum to
another, and that that variation has a long-term trend with a
characteristic timescale of the order of a century.  Included in this
variation is the Maunder Minimum, which dated from about 1645 to 1715, and indeed
there is proxy evidence, such as from tree-ring analysis, that there
were earlier similar minima, now called grand minima: the last was 
from about 1450 to 1550, and is called the Sp\"orer Minimum, before which was
the Wolf Minimum from 1280 to 1350, the Oort Minimum from 1010 to
1050, and presumably many others earlier.  The mean duration of those
minima was about 70 years, with standard deviation 25 years.  They
have occurred roughly every two and a half centuries, with standard
deviation one century.  It seems, therefore, that we are now due for
another.

What determines the sunspot-cycle period?  Or perhaps one should ask
more appropriately: what determines the period of the 22-year magnetic
cycle?  Perhaps the first idea to be put forward was by C. Wal\'en,
who suggested that the cycle is essentially a manifestation of a
magnetic oscillation of the entire Sun.  One can easily estimate the
intensity of a global magnetic field required to produce an
oscillation with a 22-year period; its precise value depends on the
geometry of the field, but all plausible geometries yield fields of
the order of 3000G, the very value observed to be present in sunspot
umbrae.  More modern ideas suppose the cycle to be determined by what
has been called dynamo action, the complicated process of field
augmentation and decay caused by magnetohydrodynamical stretching and
twisting moderated by Ohmic diffusion in and immediately beneath the
turbulent convection zone.  The 22-year cycle period does not emerge
from this scenario in so natural a manner as it does from the
global-oscillation postulate.  But it can be rationalized.  However, I
shall not attempt to describe in this lecture the panoply of theories
that have been invented to explain it, but instead refer to the
excellent recently published book on Sunspots and Starspots by Jack
Thomas and Nigel Weiss which also points the reader to more detailed
literature.

There has been much discussion about the extent to which the sunspot
cycle can be predicted.  It seems that most investigators believe that
there is a degree of predictability, the interval between, say, one
maximum and the next, being influenced by -- in the extreme view
completely determined by -- what transpired before.  This notion was
advanced some three decades ago by Bob Dicke, who noticed that the
unusually early arrivals of the 1778 and the 1788 maxima were followed
immediately by some compensating long inter-maximum intervals,
apparently trying to restore the cycle to a regular oscillation.
Others later have purveyed more complicated relations.  They all imply
that the mechanism of sunspot production has memory.

An interesting (at least to me) exercise triggered by Dicke's remark
was simply to try to answer the question: is the Sun a clock?  One can
invent two extreme, admittedly highly simplified, models.  The first
is to presume that the Sun is a clock, whose timing is controlled by a
Wal$\rm \acute{e}$n-like oscillation but whose manifestation at the
surface through sunspots has a random time lag, random because the
information about the interior must travel through the turbulent
convection zone, which occupies the outer 30 per cent, by radius, of
the Sun (see Fig.~8), yet accounts for but 2 per cent of the mass.
At the other extreme one can posit that, as dynamo theorists believe,
the cycle is controlled entirely in or immediately beneath the
convection zone where the dynamics is turbulent, and thereby, on a
timescale of 22 years, it has no memory at all.  Then the cycle period
itself is a random function.  I hasten to add that this model is
actually more extreme than most dynamo theorists accept.  The apparent
phase maintenance predicted by these two models has been compared with
sunspot data by both Dicke and myself, with similar results where our
analyses overlap; however we did not draw similar overall conclusions.
I think it is fair to say that the solar data lie between the two
extremes, suggesting that the Sun has a modicum of memory, as many
dynamo theorists would maintain.

Sunspot-cycle predictability, and with it actual prediction, has come
into vogue in recent times.  But before remarking on current
happenings, I shall relate a pertinent story which exposes an
important variance of opinion concerning scientific inference.  Nearly
four decades ago I met Charlie Barnes, the chief keeper of time at
what was then called the National Bureau of Standards, in Boulder,
Colorado, USA.  In a digression from his usual activities he had
addressed sunspot-cycle variability from the viewpoint of his
modelling the random fluctuations in precision timing by caesium
clocks.  He had a simple mathematical model, basically a filter which
in effect accepted only a part of a time series, concentrated mainly
in a given frequency band.  Thus, if one sent a random signal through
the filter, one received as output a quasi-periodic response which,
after rectification, could be compared with sunspot numbers.  The only
pertinent parameter he could adjust is the ratio of the width of the
filter to its central frequency.  Barnes calibrated that ratio, first
by requiring that the variance of the cycle period was the same as
that of the sunspot number, and then by requiring that the variance of
the heights of the maxima agreed with the variance of the sunspot
numbers at maximum.  The two calibrations gave the same result.
Barnes then pointed out that if one ran the model backwards the
original random signal (save for a component that does not influence
the output) was recovered, because the whole (linear) process was
determinate in both directions.  So one could run the machinery
backwards feeding it with the actual sunspot data, obtaining an
apparently random result, and then run it forwards to recover the
original data.  What Barnes knew is that if one ran it forwards and,
at some moment, stopped the input, the output is the most likely
outcome of the process.  He therefore had a predicting machine, which
he had tested by truncating the apparently random input early, and
seeing how well his mathematical machinery \textquoteleft predicted'
what should follow.  It performed rather well.  I was so excited by
this result that I went straight up the hill to the High Altitude
Observatory, which in those days was situated on a mesa above the
National Bureau of Standards at the National Center for Atmospheric
Research.  There I encountered Peter Gilman, and enthusiastically
described to him this fascinating result.  \textquoteleft It has no
interest whatever', retorted Gilman, \textquoteleft because it
contains no physics'.  But I disagreed strongly, for it is indeed
extremely interesting, and the reason for it being so interesting is
because it apparently contains no physics; if one wishes to
demonstrate the validity of the physics that has been put into a
theory by comparing its consequences (I refrain from calling them
predictions because so often these consequences are post hoc) with
observation, one must surely demonstrate that one has done
significantly better than a physics-free procedure\footnote{or one
  must demonstrate that the physics-free procedure happens, by chance,
  to model the physics of the process under investigation.}.

I now come to real prediction.  Or shall I call it sociology?
Currently there are (at least) two identical games being played --
competitions in waiting whereby scientists have deposited with
adjudicators their estimates of the sunspot number at the next
maximum.  It is supposed to be a bit of harmless fun.  I should stress
that fun is scientifically useful, a view with which I am sure Vainu
Bappu would agree, for it provides rejuvenating relief from the
serious pursuit of discovery which occupies most of our lives.  But
what will the reaction be when the results of the competitions are
known?  Will the winners claim that the theories they have used are
vindicated?  Although the entries have been kept confidential by the
adjudicators, I do know from talking to some of the competitors that
there is substantial diversity amongst the procedures that have been
adopted for determining them, procedures which at some level are
presumably being tested.  One can imagine, for example, that Gilman
and his colleague Matsumi Dikpati, who have made much of their ability
to predict the solar cycle, will have entered, hoping, perhaps, to
vindicate their theory.  Their model requires several parameters to be
calibrated, so one should heed Pauli's warning.  There are also purely
mathematical, less deterministic, algorithms which in a
less-easily-appreciated manner incorporate history into a statistical
foretelling.  At the other extreme, Weiss and David Hughes, for
example, believe that the cycle is inherently chaotic, albeit with an
underlying control which, turbulent convection aside, is
deterministic.  Therefore any prediction must be very uncertain.  What
might either of them have submitted, if indeed they have entered the
fray?  There is a diversity too amongst the reasons for entering the
competition.  I have entered one of the competitions myself, but I
shall keep quiet about my motives until the matter is settled.  One
thing we do know is that there are many competitors, with entries that
must surely range from near zero, submitted by those who believe that
we are plunging into the next grand minimum (at the time writing there
are many fewer sunspots than most spectators have expected) to values
comparable with the highest ever recorded.  Therefore the range of
possibilities is bound to be densely sampled, as would have been the
case had everyone submitted random numbers.  So the winners are
therefore bound to be very close to the actual result.

\section{What causes Sunspot Darkening?}

It's the magnetic field.  That field can roughly be thought of as an
ensemble of elastic bands imbedded in the fluid, such as the flux
tubes illustrated in Fig.~9.
  
Before embarking on a discussion of the physics of sunspots, I must
point out what is actually meant by the term `sunspot'.  As was
evident in my introduction, initially the term was considered to
denote simply a dark patch on the Sun's surface like those illustrated
by Figs.~4 and 10.  But now it is considered also to be the entire
three-dimensional edifice, extending upwards from the dynamically
controlling layers beneath the photosphere into the consequent
magnetically active region above it in the atmosphere.  I shall use
the term in both senses, I hope without ambiguity.

\begin{figure}  
\begin{center}
  \includegraphics[width=0.6\textwidth]{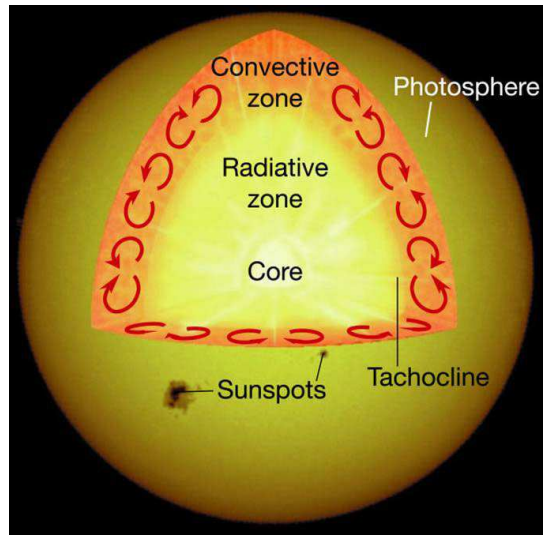}
\caption[]{Simple representation of the Sun, showing in a cut-out the major
zones.  The curved arrows represent convective overturning.}
\end{center}
\end{figure}

Magnetic field resists being stretched, and therefore opposes any
shear in the fluid that would induce stretching -- in other words, it
reacts against a fluid velocity with a transverse component that
varies in the direction of the field.  The energy generated by nuclear
reactions in the core of the Sun is transported outwards through the
majority of the surrounding envelope by photon diffusion, but in the
outer 30 per cent (by radius) the fluid is buoyantly unstable, and the
energy is carried almost entirely by convection, which consists of
overturning eddies (illustrated by curved arrows in Fig.~8).  The
magnetic field hinders the overturning, and in the umbra is strong
enough to stop the normal convection entirely, at least in the very
upper layers of the convection zone where the fluid density (inertia)
is relatively low and is incapable of overcoming the stresses imposed
by the magnetic field.  Some motion can occur, however; it provides a
weak vehicle to transport energy, and is responsible for the umbral
structure observed in the photosphere, but that is of secondary
importance to the broad overall picture I am painting here.

\begin{figure}  
\begin{center}
  \includegraphics[width=0.8\textwidth]{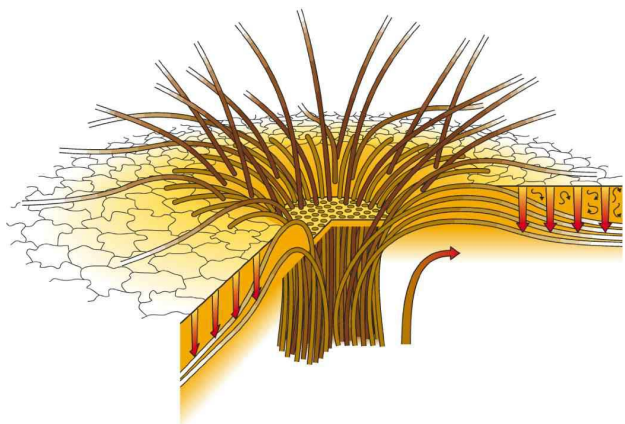}
  \caption[]{Sketch by Weiss, Thomas, Brummell and Tobias of the upper
    portion of a sunspot.  The magnetic field is held together,
    presumably by converging fluid beneath the region illustrated, in
    a vertical umbral column, and then splays out through and above
    the penumbra where the fluid is unable to confine it, alternating
    between flux tubes rising almost freely into the upper atmosphere
    and tubes forced back beneath the photosphere by descending
    convective flow (indicated by the broad vertical arrows).}
\end{center}
\end{figure}

The geometry of the field is illustrated by the tube-like structures,
sometimes called ropes, in Fig.~9, drawn by Weiss, Thomas, Nic
Brummell and Steve Tobias.  The tubes are concentrated in a vertical
column underneath the umbra.  Some care should be exercised in
interpreting the illustration, which should not be taken too
literally.  It gives the impression that the umbral field is contained
in the tubes with little or no field between them.  That is almost
certainly not the case; instead the field is bound to be much smoother
on the transverse scale of the tubes.  It should be appreciated also
that the orientation of the field is the same in all the ropes.

I must now point out another property of magnetic fields: not only do
they exert a tension along the tubes, endowing the fluid with a degree
of elasticity, but they also exert a transverse pressure --
neighbouring field tubes of the same polarity (magnetic fields
parallel) repel one another; conversely, tubes of opposite polarity
(magnetic fields antiparallel) apparently attract, and annihilate each
other, dissipating much of their energy into heat and converting the
rest into kinetic energy of the fluid. This process, generically
called reconnection, is very complicated, and is an arena of very
active research.  It is of particular interest in the atmosphere above
and near sunspots, where the activity is visible.  It is no doubt just
as important, if not more so, beneath the photosphere where it cannot
be seen, and where either the fluid is less of a slave to the field or
the field is a slave to the fluid.  But I digress.  Returning to the
concentrated umbral field, it is evident that there must be some force
holding the field together.  The only possibility is an inward
(towards the \textquoteleft axis' of the spot) momentum flux carried
by fluid converging at depths where its inertia is great enough to
dominate the dynamics.  Near the surface of the Sun the fluid can no
longer contain the field; the field splays out, becoming weaker and
more nearly horizontal.  It can no longer prevent the convection
(mainly because of the changed orientation), which tends to obviate
field stretching by forming elongated eddies, aligned with the field,
whose motion is predominantly transverse to the field, producing the
penumbral filaments.  Moreover, the surrounding fluid no longer
converges on the spot, but diverges, at least in places, as was
observed by Evershed a hundred years ago.

In the picture provided by Weiss and his colleagues, which is based on
prior superficial observation, the field does not splay out smoothly
into the penumbra; instead there is an alternation of gradually
splaying flux tubes which extend high into the atmosphere and more
nearly horizontal tubes that tip back below the photosphere near the
edge of the penumbra, pushed down, it is believed, by granular
convective motion that is not seriously impeded by magnetic field and
which has an up-down asymmetry of such a nature that descending fluid
has the greater influence on the magnetic field.  That process is
called magnetic pumping, and is represented by the downward arrows in
the figure.  It holds the field down against both the natural tendency
of the field to want to be straight (because of its tension) and
against buoyancy: magnetic field exerts transverse pressure which
equilibrates with the pressure in the surrounding fluid, the fluid
requiring density (inertia, and therefore gravitational mass) to exert
pressure, whereas the field has none; regions of concentrated field
are less dense than their immediate surroundings and are therefore
buoyant.  In the inner penumbra where the inclinations of the
alternating magnetic flux tubes do not differ greatly, the elongated
rolls raise the field where the hot bright fluid ascends and depress
it where the cool darker fluid descends.  Further out where the
inclinations differ substantially, the interaction between the motion in the bright filaments and that in the dark horizontal
filaments is probably weaker.  It is along the near-horizontal darker tubes that the
Evershed motion is driven by a pressure gradient that is insufficient
to push fluid high into the atmosphere along the more inclined (from
the horizontal) field.  What produces that pressure gradient appears
not to be well understood.  I should point out that other scenarios
have been suggested in the literature; once again, I refer the reader
to Thomas and Weiss's book for details.

I come back now to the question posed by the title of this section.
Except in a very thin superadiabatic boundary layer at the top of the
convection zone, almost all the heat from the nuclear reactions in the
core is transported through the convection zone by material motion.
As I have already indicated, that transport is inhibited in a sunspot
by the magnetic field.  Therefore less heat gets through, one might
naturally think, and the spot must obviously be dark.  That conclusion
is basically correct, although with a little more thought one must
realize that it is actually not entirely obvious.  It depends on
certain conditions being satisfied, namely that the spot is a small
superficial blemish on a deep convection zone -- and by small I mean
having both a lateral lengthscale and a depth that are much less than
the depth of the convection zone.

A spot is normally considered to have ceased to exist once a depth is
reached beyond which significant convective inhibition is no longer in
operation.  How that comes about depends on the field configuration,
which we do not know.  But we could consider two extremes.  If the
field were to extend downwards as a uniform monolithic tube, the
stress it would exert would be essentially independent of depth; gas
pressure increases monotonically downwards, however, and there must be
a level beneath which it overwhelms the magnetic stress, rendering the
field incapable of preventing convection.  In the opposite extreme, if
the field stress were to remain, say, a constant proportion of the gas
pressure -- I should point out that stress is proportional to the
square of the field strength $B$, and that the magnetic flux, which is
the product of $B$ and the cross-sectional area $\sigma$ of a magnetic
flux tube, is invariant along the tube -- then the area $\sigma$ of
the region in which the field is contained (whether it remains a
monolith or splits into spaghetti, as some investigators have
maintained), and in which there is no convection, becomes so tiny at
great depths that its presence is irrelevant to the overall picture.

The spot dams up heat beneath it, which nevertheless can readily be
transported sideways and upwards around the spot by the highly
efficacious convection without substantial modification to the
stratification in the surrounds.  There is now less heat demanding to
be carried through the spot.  The flux radiated from the surface of
the spot is less than that elsewhere, and therefore the spot is
darker; moreover the surface temperature is lower than that of the
normal photosphere, because total radiant flux is proportional to a
positive (actually the fourth) power of temperature.  With the
reduction in temperature in the spot is a consequent reduction in
pressure, which causes the material in the spot to sink under gravity
(recall that the magnetic field is essentially vertical and the field
exerts no longitudinal pressure); that is basically the reason for the
Wilson depression.  The reduction in pressure is compensated by a
lateral pressure-like stress in the horizontal from the magnetic
field, enabling the spot to be in pressure equilibrium with the
surrounding hotter, more distended, material.  Given this apparently
straightforward description one might expect spots not to be a
phenomenon associated with only the Sun.  Indeed, the presence of dark
spots has been inferred from observations of other cool stars having
deep convection zones.

The situation is not the same in hot stars.  There is overwhelming
evidence for spots on Ap stars, for example.  Indeed, both magnetic
field concentrations and coincident patches of anomalous chemical
abundance have been mapped by Doppler imaging.  But there is no
evident variation in total brightness.  (I hasten to add that some
such stars exhibit brightness variation in limited optical wavelength
bands, but that is due mainly to optical spectrum changes caused by
the abundance anomalies, and is not necessarily indicative of total
flux variation.)  The reason is that these stars have very thin
convection zones, and convection is suppressed by the magnetic field
in the spot all the way from the top to the bottom of the zone; also
the spots are very much larger than those in the Sun, having areas
which are a substantial fraction of the total area of the stellar
surface, therefore having a linear lateral dimension which is very
much greater than the depth of the convection zone.  Heat cannot
easily escape around the edges of the spot by flowing laterally great
distances though the ill-conducting radiative zone beneath.  Instead,
the stratification in the spot is forced to adjust to accommodate the
heat flux demanded by the radiative interior.  That adjustment is one
in which the spot region becomes more distended, noticeably so if one
measures the distension in units of the convection-zone depth, but by
only a very small amount relative to the total radius of the star:
there is what one might call a Wilson elevation.

I should point out that these two descriptions of spots do not
encompass all possibilities: there are also stars whose structure is
intermediate between that of the Sun and those of what I have called
hot stars; they also support spots, and those spots produce some
genuine local diminution of the total radiative flux.  Why have I
digressed so far from the Sun to describe a situation which is hardly
relevant to sunspots?  The reason is simply to stress that the physics
of sunspots is more subtle than one might have first suspected, and
that suppression of the {\it mechanism} of heat transport in a star
does not necessarily result in substantial suppression of the {\it
  amount} of heat that is transported.

\begin{figure}  
\begin{center}
  \includegraphics[width=0.8\textwidth]{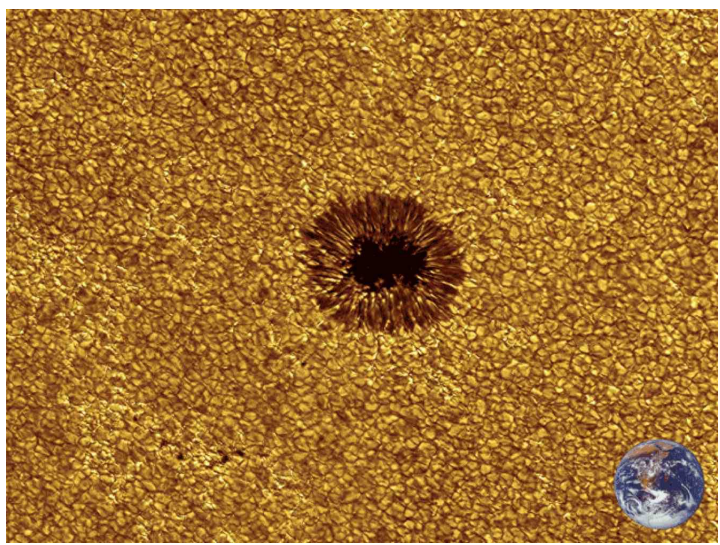}
  \caption[]{G-band image of a portion of the surface of the Sun
    containing a medium-sized sunspot.  Superposed in the bottom
    right-hand corner is a image of the Earth, to provide a graphic
    comparison of scale.  The mean intensity of the surrounding
    convective flow, the solar granulation, appears to vary on a scale
    much larger than the granules, but in patches that are apparently
    random, with no obvious bright ring around the spot.}
\end{center}
\end{figure}

The process of diverting the heat around a sunspot was first
considered seriously by Henk Spruit.  The motivation for his study was
that others had speculated earlier that the missing heat flux should
be radiated from a necessarily bright annulus around the spot of
thickness comparable with the spot's radius, but that the brightening
had not been observed (see Fig.~10).  In his study Spruit assumed
the convective motion to be everywhere on a scale much smaller than the
scale of variation of the heat flow, and he ignored the presence of
any large-scale flow induced by the disturbance to the temperature
variation produced by the suppression of the convective heat transport
in the spot.  He also ignored the effect of the large-scale
temperature disturbance on the convection, so that the heat transport
could be described as simply a classical diffusive process with a
temporally unvarying diffusion coefficient, the value of which Spruit
obtained from mixing-length theory.  Spruit considered the evolution
of the temperature distribution after suddenly imposing a heat plug in
the outer layers of the convection zone to represent the creation of a
sunspot.  He confirmed a view that was already held by some, although
perhaps it had not been well substantiated, that because the turbulent
diffusion coefficient and the heat capacity of the convection zone are
both so high, transport around the spot is facile and extensive: most
of the heat blocked by the spot is distributed throughout the
convection zone, almost all of which could easily be retained over the
lifetime of a spot (the cooling time of the convection zone is 10$^5$
years), and that which is radiated around the spot is distributed so
widely that its influence on the photosphere is undetectable, in
agreement with observation.  It should perhaps be commented that the
calculation is highly idealized, even in the context of mixing-length
theory.  The speed of propagation of the greater part of the thermal
disturbance produced by the introduction of the plug is comparable
with the convective velocities, which invalidates the diffusion
equation that was used: purely thermal disturbances cannot travel
faster than the convective motion that advects them (admittedly the
associated \textquoteleft hydrostatic' readjustment is transmitted at
the speed of sound, but the magnitude of the large-scale adjustment is
tiny), which is contrary to the formally infinite speed permitted by a
classical diffusion equation.  Instead the transport equation should
have a wave-like component, somewhat analogous to the telegraph
equation.  Moreover, temperature fluctuations are not passive, but
influence the buoyancy force that drives the very convection that
transports them.  That back reaction modifies the wave-like term in
the transport equation.  Nevertheless, because the convection zone is
so close to being adiabatically stratified (except in a thin boundary
layer), these niceties play little role in the overall structure of
the Sun, and Spruit's basic conclusions must surely be right.

\section{The Rotation of the Sun}

I have already remarked that in the early days Galileo, Fabricius,
Scheiner and others had inferred from the motion of sunspots across
the disc that the Sun rotates.  Subsequent observations have mapped
the angular velocity in greater detail, and in modern times those
results have been broadly confirmed by direct Doppler observations of
the photospheric layers; the different measures are not precisely the
same, but that is because Doppler observations see only the surface of
the Sun whereas sunspots extend below the surface and presumably
rotate with some average over their depth, which we now know is not
quite the same.  Nevertheless, the basic picture is one of a smooth
decline in rotation rate from equator to pole, the rotation period
(viewed from an inertial frame of reference, not rotating with the
Earth) increasing from about 25.4 days at the equator to something
like 36 days at the poles; the latter value is only approximate
because it is difficult to view the poles (recall that the axis of solar rotation
is inclined by only 7$\rm ^o$ from the normal to the plane of the
ecliptic), and, of course, sunspot motion itself cannot be measured
because sunspots are found only equatorward of latitudes $\pm 30\rm
^o$ or so, so other indicators have had to be followed.

Rotation well beneath the surface has only recently been measured, by
seismology with acoustic waves.  I shall describe briefly how that is
done.  Acoustic waves are generated essentially as noise by the
turbulence in the convection zone, and reverberate around the Sun.
Any given wave propagates around the Sun, confined (approximately) to
a plane, as illustrated in Fig.~11.  They are reflected near the
surface of the Sun, typically somewhat below the upper superadiabatic
boundary layer of the convection zone where the scale of variation of
the density and pressure is comparable with or less than the inverse
wavenumber of the waves, thereby preventing those waves from
propagating upwards into the atmosphere -- the condition for
propagation of an acoustic wave to be possible is that, roughly
speaking, the scale height of the background state must exceed
1/4$\pi$ of the wavelength of the wave.  Downwardly propagating waves
are refracted back towards the surface by the rising sound speed
caused mainly by the increase of temperature with depth.  Therefore
waves of a given inclination are trapped in an annulus, whose inner
boundary is represented by the dotted circles in the figure.  (I am
assuming for the purposes of the introduction to this discussion that
the Sun is basically spherically symmetric), and their properties are
determined by conditions in that shell: the relation between the wave
frequency and the observable wavenumber at the surface is an indicator
of average conditions in the shell, the average being weighted by a
function proportional to the time spent by the wave in any particular
region.  Segments of four sample ray paths (essentially the paths
followed by the waves) of differently directed waves are illustrated
in Fig.~11; there are other paths, similar to those illustrated,
lying in planes through the centre of the Sun but inclined to the one
illustrated -- for example, out of the page towards us at the top and
away from us at the bottom, or vice versa.

\begin{figure}  
\begin{center}
  \includegraphics[width=0.6\textwidth]{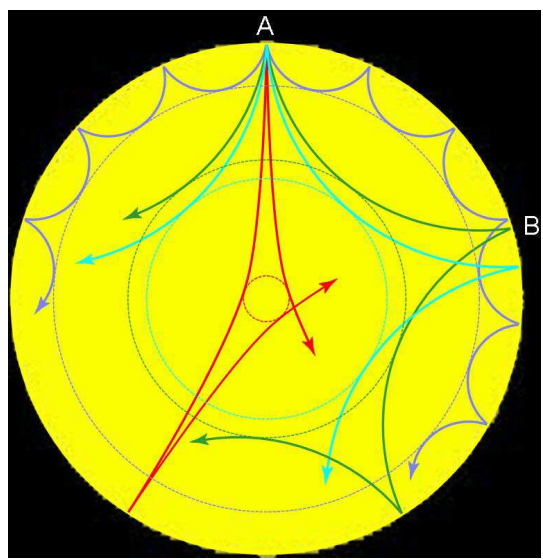}
  \caption[]{Segments of ray paths followed by acoustic waves in the
    Sun.  The dotted circles represent the envelopes of the lower
    turning points (lowest points of the ray paths) of the waves.}
\end{center}
\end{figure}

The essence of the procedure for mapping the solar interior is as
follows: Suppose we were to know the wave speed in the Sun down to the
bottom of the shell containing, say, the second most deeply
penetrating wave illustrated in the figure.  Then we can actually
calculate the properties of that wave, and also that of the first,
shallowest, wave, and, indeed, of all other waves that are shallower
than our selected second wave.  Consider now the third wave, which
penetrates only slightly more deeply than the second.  Evidently we
could calculate its progress throughout most of its passage; what is
missing is the almost horizontal passage through the very thin annulus
occupying the space between its deepest penetration level and that of
the second wave: the space between the second and third dotted circles
in Fig.~11. We can therefore represent the observable properties of
that wave -- in particular the relation between its frequency and its
horizontal wavenumber at the surface of the Sun -- in terms of the
average wave speed, I call it $\bar{c}$, in that thin annulus.
Measurement of the surface wavenumber and frequency then provides the
essential datum to determine $\bar{c}$.  We have thereby extended our
knowledge of the wave speed down to a lower level.  By considering
successively more and more deeply penetrating waves we can, provided
we have observations of a sufficient range of waves, build up a
somewhat blurred view of the wave speed throughout the entire Sun, the
blurring being because we are actually measuring averages over the
annuli between adjacent lower boundaries of different regions of wave
propagation, not point values.  One can then combine with that
information corresponding results from similar sets of waves
propagating in planes inclined to the first, and thereby in principle
build up a three-dimensional picture of the wave speed throughout the
Sun.

An obvious apparent flaw in my argument is that if all the waves are
reflected beneath, rather than at, the surface of the Sun, one cannot
know the structure of the Sun all the way to the surface.  So how can
one proceed?  And how can the trapped waves even be observed at the
surface?  The answer to the second question is that even though the
motion at the upper reflecting boundary of the region of propagation
cannot formally propagate to the surface, the surface layers do
respond as a whole to that motion, being simply lifted up and down in
approximate synchronism with the wave below.  (I admit to speaking
rather loosely here, but as a first approximation it is safe to regard
that statement as being true.)  Therefore the wave motion below is
observable.  Its influence on the motion of the photosphere is
portrayed by the Doppler images in Fig.~12.  One can now address the
first question by simply representing the surface layers by some
average impedance, much as we represented the wave speed between the
lower boundaries of the regions of propagation of the second and third
waves by an appropriate average $\bar{c}$.  Fortunately the upper
boundaries of the regions of propagation of all the waves are roughly
in the same place, so the impedance for all waves does not vary a
great deal.  (The range of observable frequencies, roughly 2 - 4 mHz,
which also influence -- fortunately only weakly -- the impedence
somewhat, is not great.)  This represents a fundamental uncertainty in
the inferences, but that uncertainty becomes smaller and smaller the
deeper in the star one's inferences are drawn.

\begin{figure}  
\begin{center}
  \includegraphics[height=4.7cm]{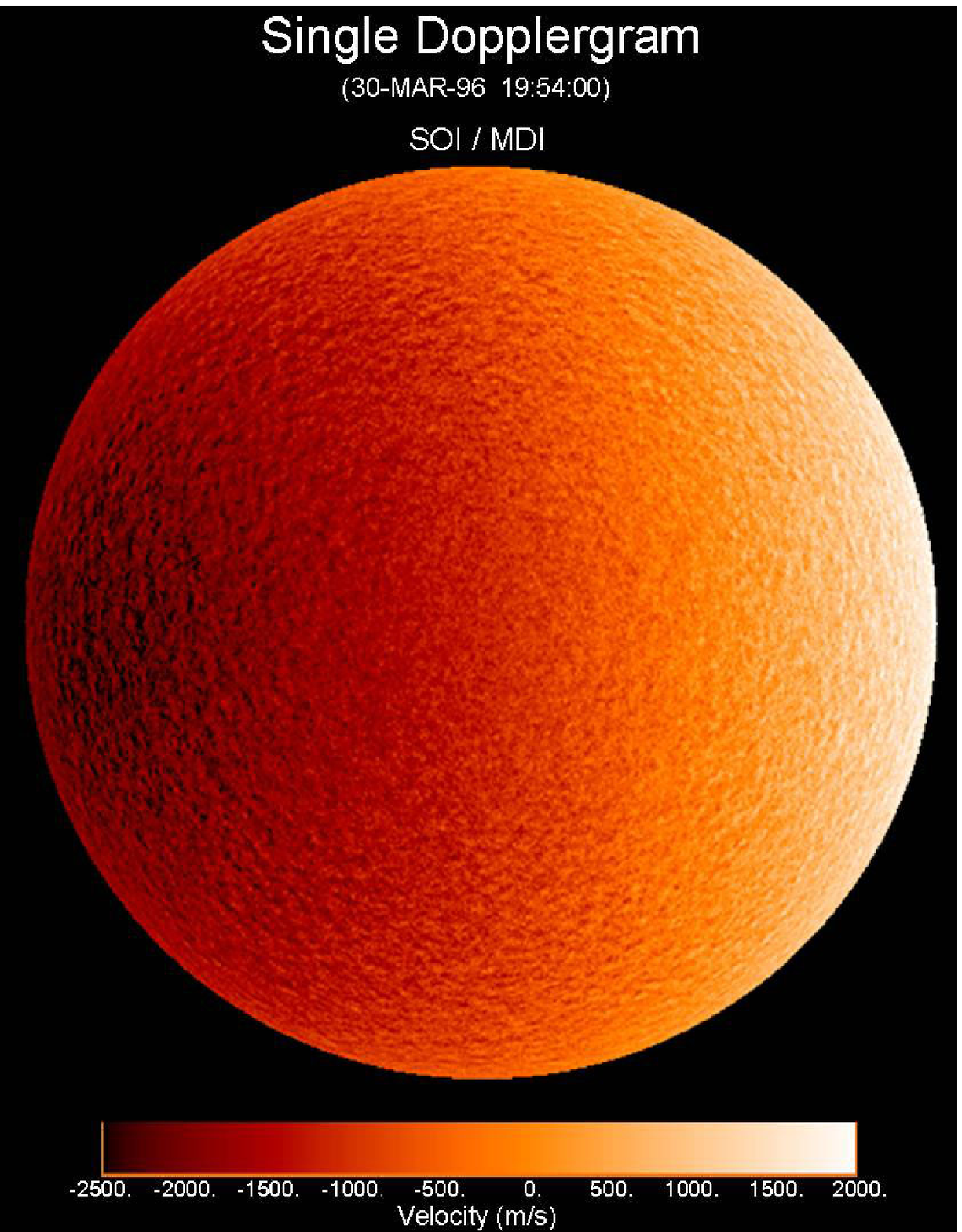}
  \includegraphics[height=4.7cm]{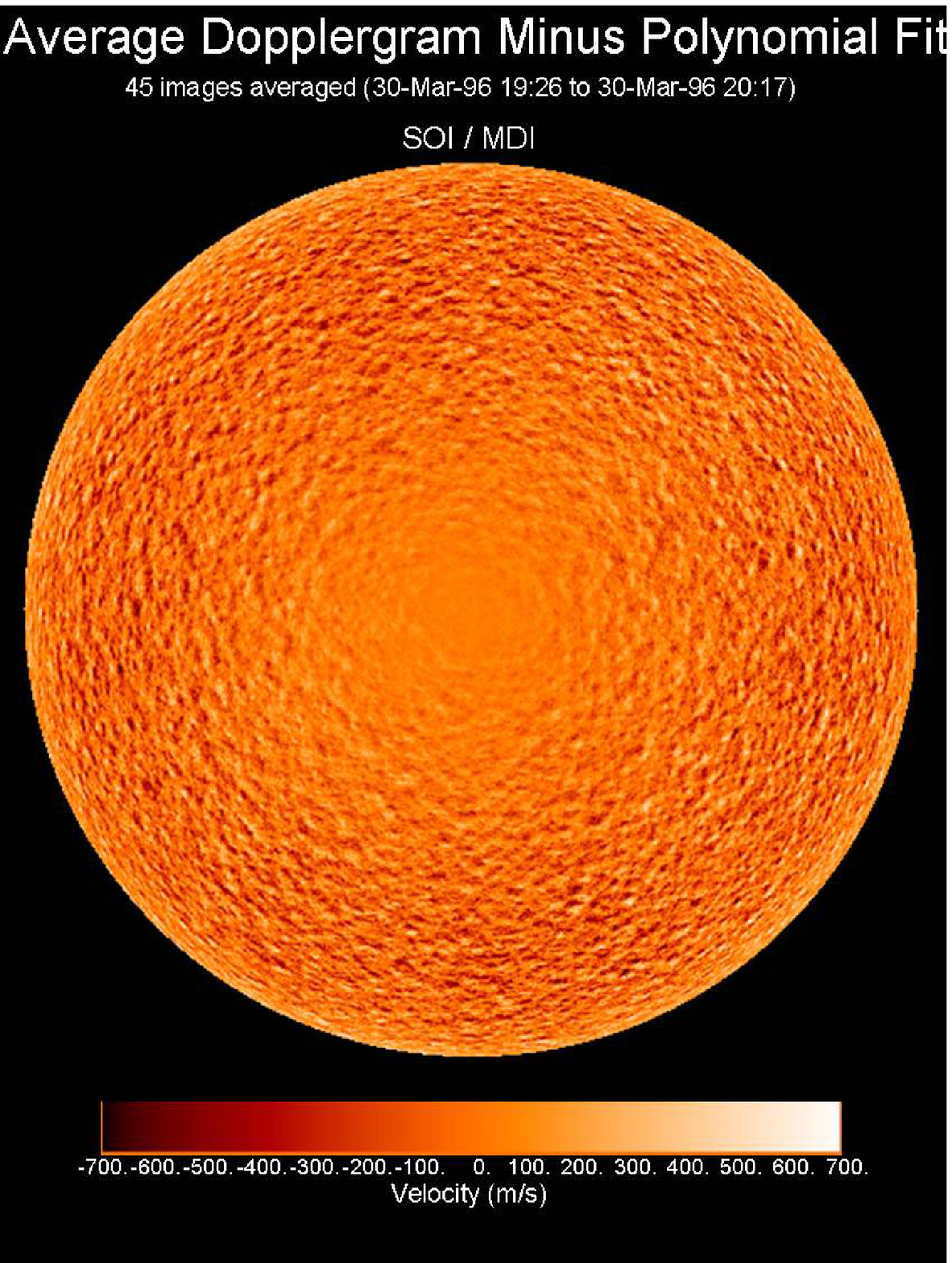}
  \includegraphics[height=4.7cm]{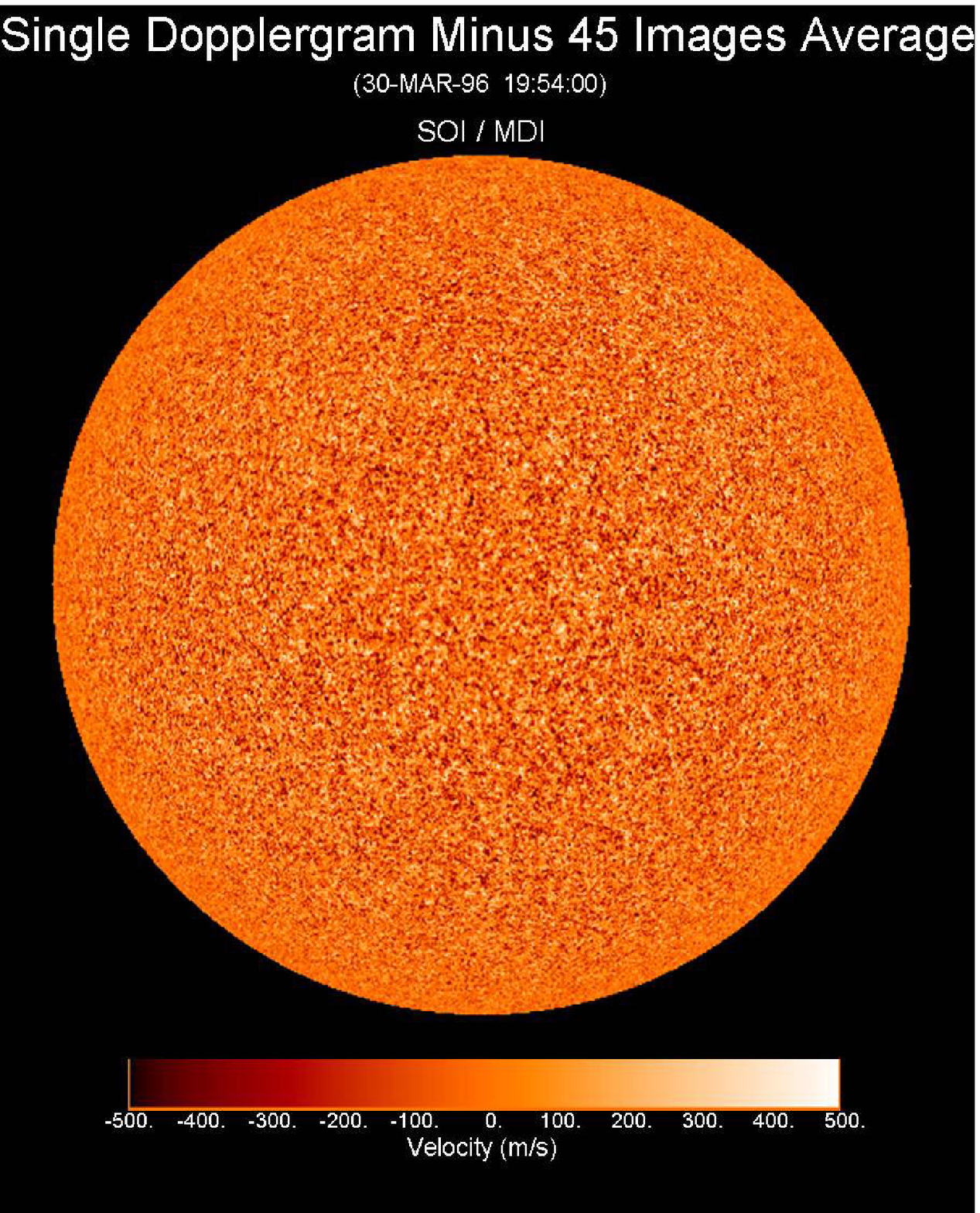}
  \caption[]{Doppler images of the Sun, obtained by the Solar
    Oscillations Investigation using the Michelson Doppler Imager on
    the spacecraft SoHO.  Dark shading represents line-of-sight
    velocity towards the observer, light represents velocity away.
    The values of the velocities represented by the greyscales are
    indicated at the bottom of each panel. The first panel is a raw
    Dopplergram; it is dominated by the Sun's rotation, although
    superposed smaller-scale motion is evident.  The second panel is
    an average of 45 images (which suppresses the oscillations and
    granular convective motion, although the resolution is inadequate
    to resolve granules) from which the contribution from rotation has
    been subtracted; what is left are the tops of the supergranular
    convective cells, whose velocities are more-or-less horizontal,
    and therefore is most visible towards the limb (although not too
    close where foreshortening is severe), and invisible at disc
    centre.  The third panel is a single Dopplergram from which the
    45-image average has been subtracted, thereby removing rotation
    and supergranulation, leaving principally the acoustic
    oscillations, whose velocity in the photosphere is almost
    vertical; the amplitude observed is therefore greatest at disc
    centre.  Notice that the magnitudes of the oscillation velocities
    are comparable with the convective velocities, approximately
    $0.5\, {\rm km\, s}^{-1}$.  For comparison, the sound speed in the
    photosphere is about $7\, {\rm km\, s}^{-1}$.  The sound speed at a
    level near the base of the sunspot (say, 7\,Mm) is about $30\,
    {\rm km\, s}^{-1}$.  }
\end{center}
\end{figure}

Let me now address what we can deduce from the wave-speed inferences.
In the absence of a significant magnetic field the wave speed relative
to the fluid is essentially a local property of the fluid; it is
dominated by what we normally call the sound speed, which depends just
on pressure and density (and somewhat on chemical composition), but is
modified a little by stratification.  In addition the wave is
\textquoteleft carried' by the fluid motion, the latter being mainly a
consequence of the rotation of the Sun.  So one can measure the
wave-speed averages in the manner I have just described, first from a
set of waves all of which have an eastward component of propagation,
and then from a similar set of waves with a westward component.  Their
average is then the intrinsic wave speed, relative to the fluid, and
their difference is twice the rotation velocity of the Sun.  Much
physics has been learned from the intrinsic wave speed, because it is
directly related to the properties of the material of which the Sun is
composed, at least in regions where magnetic stresses are negligible.
But that is not the subject of this lecture.  Instead I shall comment
briefly just on the rotation.

\begin{figure}  
\begin{center}
  \includegraphics[width=0.6\textwidth]{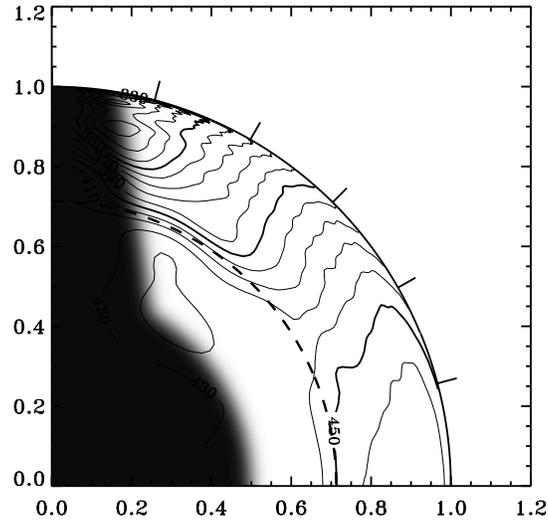}
  \caption[]{Contours of constant angular velocity in the Sun.  The
    blacked-out regions mark where it has not been possible to draw
    reliable inferences (from a study by Jesper Schou and his many
    participating colleagues).}
\end{center}
\end{figure}

The rotation rate in a quadrant of the Sun is depicted in Fig.~13.
Plotted are contours of constant rotation rate.  Adjacent contours are
separated by 10~nHz.  The method used to construct this diagram
produces only an average of the rotation in the northern and southern
hemispheres, which is why only a quadrant is displayed.  It is evident
that, broadly speaking, the latitudinal variation of the rotation that
had been observed at the surface persists with only minor change right
through the convection zone.  But the radiative zone rotates
uniformly.  There is a thin shearing layer at the base of the
convection zone, called the tachocline, which is too thin to be
resolved.  It is here that many dynamo theorists believe that magnetic
field is augmented and, temporarily, stored, producing the solar
cycle.  I have already promised not to discuss the details.  One
feature of the plot to which I would like to draw attention, however,
is that the shear, and therefore any consequent stretching and winding
of the (dynamically weak) magnetic field that might be present,
reverses direction at a latitude of about 30$\rm^o$.  That is just the
latitude at which sunspots first form at the beginning of each new
solar cycle (Fig.~6).  Surely that must provide a clue to the
mechanism of the cycle.  Or is it mere fortuitous coincidence?

\section{The overall structure of a large sunspot}

Only the larger sunspots have a nice well defined structure with
surface appearance like those illustrated in Figs.~4 and 10.  Small
spots contain less magnetic flux and are less able to control the
turbulent convective flow in which they are imbedded.  They are
consequently much less regular.  I shall therefore confine my
discussion to the relatively clear prototypical case, thereby avoiding
having to describe the gamut of smaller magnetic structures that are
visible on the surface of the Sun: if I were to do otherwise this
lecture may never end.

\begin{figure}  
\begin{center}
  \includegraphics[width=0.5\textwidth]{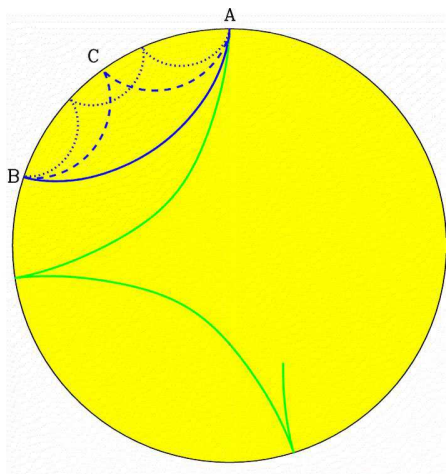}
  \caption[]{Sketch of sample ray paths adopted by Zhao, Kosovichev
    and Duvall for inferring sunspot structure.  Points like A and B
    mark typical observation points in the quiet Sun; point D is the
    location of a sunspot.  The distances between A, B and C, and
    hence the depths of the ray paths joining them, have been
    exaggerated for clarity.  The continuous ray paths are typical
    spot-free paths, like those depicted in Fig.~11, from which the
    background (spot-free) structure is inferred.}
\end{center}
\end{figure}

The properties of a large sunspot and its immediate surrounds have
been mapped by acoustic seismology by Jun Wei Zhao, Sasha Kosovichev
and Tom Duvall.  To a large extent they are consistent with the
picture I have been building up during this lecture, although one
essential ingredient is missing, namely the Evershed flow.  In
principle, the method of inference that was employed to obtain this
picture is much as I described for determining the Sun's rotation; the
difference is just in the detail, which is a little more complicated.
Consider the three ray-path segments joining observation points A and
B in Fig.~14; the point C marks the location of a sunspot.  The
continuous ray paths are examples from the set considered in \S6, and
are drawn simply as a benchmark; they are unperturbed by the shallow
sunspot.  The dotted ray path passes underneath the sunspot and may
feel some influence from it, and the dashed path evidently passes
through the spot.  By comparing observed propagation times from A to B
and from B to A of the dotted and dashed waves with those of similar
wave segments in another location where there is no sunspot, the
influence of the sunspot can be ascertained.  As always, the answer is
a new average propagation speed $\bar{c}$ along the ray paths.  One
must then tackle the complicated geometrical problem of unravelling
those averages over a wide variety of rays to obtain genuinely
localized averages, of both intrinsic propagation speed and of fluid
flow, for such averages are comprehended more easily than the raw
ray-path averages.  I shall not go into the details of how the
unravelling is accomplished; for the purposes of the present
discussion it is adequate to consider the task to be just a
technicality which we know how to handle.

The outcome is illustrated in Fig.~15.  What is shown is a section
in a rotatable vertical plane of a three-dimensional representation of
a measure of the intrinsic wave propagation speed and the large-scale
fluid flow -- only a single orientation of the plane is illustrated in
the figure reproduced here.  The shading represents the intrinsic wave
speed and the arrows represent the flow, their size denoting the
magnitude of the velocity.  The intrinsic wave speed is difficult to
interpret: it is influenced by both the temperature of the fluid and
the magnetic field, which the current measurements cannot disentangle;
even more uncertainty is added by the fact that the effect of the
magnetic field is anisotropic, being a more-or-less increasing
function of the inclination of the direction of wave propagation to
the direction of the field -- what is illustrated by shading in the
figure is only a scalar, presumably an average over the particular
waves that have been used for the inference, weighted by the relative
importance that the localization procedure adopted by the analysis has
given to those waves.  Interpretation must therefore entail some
guesswork.  It is likely that the wave speed illustrated in the figure
is due predominantly to temperature, because immediately beneath the
photosphere both field and acoustic wave propagation are both very
nearly vertical, and consequently parallel to each other, and
therefore hardly interact.  Moreover, as I have already described, at
depth the influence of the field declines dramatically either because,
unlike the gas pressure, the intensity of the field does not increase
significantly with depth, or because the proportion of the volume
occupied by the field diminishes greatly.  (It is worth pointing out
that because the lateral field stress under the umbra balances the gas
pressure deficit produced by the lowering of the temperature, a
putative horizontally propagating acoustic wave would be influenced by
comparable amounts, although oppositely, by field and negative
temperature change.  Those influences would not exactly cancel,
however, because the effective adiabatic compressibilities of field
and gas, which control the wave speed, are different.)  Therefore I
may lapse into \textquoteleft hotter' and \textquoteleft colder' as a
convenient device to describe wave-propagation-speed differences
succinctly.

\begin{figure}  
\begin{center}
  \includegraphics[width=0.8\textwidth]{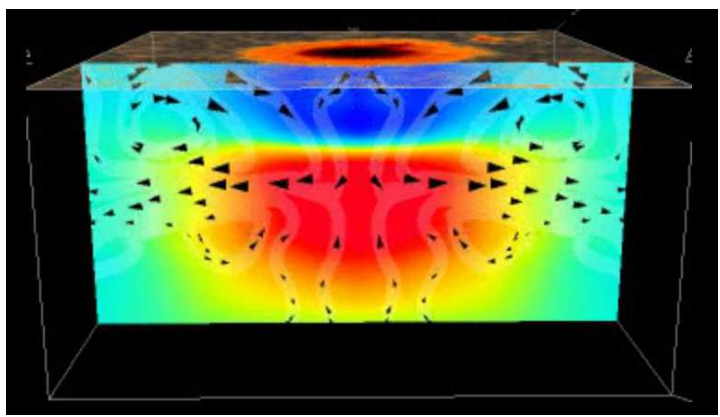}
  \caption[]{Seismic image of a sunspot by Zhao, Kosovichev and
    Duvall.  The shading represents the deviation of the wave
    propagation speed from that in a corresponding spot-free region of
    the Sun, dark (in this black-and-white picture) denoting both
    positive and negatives values.  The arrowheads indicate the
    direction and magnitude (denoted by their size) of the flow.}
\end{center}
\end{figure}

The dark shading in Fig.~15 immediately beneath the upper surface of
the spot is to be expected: the surface of the spot is cool, and, as I
have already explained, so should be the underlying fluid where
convection is suppressed by the magnetic field.  There is a second
relatively dark region lower down in this black-and-white image, this
time representing hotter fluid, presumably beneath the region in which
convection is suppressed -- in other words, beneath the spot.  This is
where heat from below is dammed up, being unable to pass easily
through the spot.  In a broad sense, the fluid flow associated with
these temperature (actually wave-speed) anomalies is easy to
understand -- at least it seems superficially to be that way.  The
cool plug beneath the surface cools the surrounding fluid, causing it
to sink in a negatively buoyant cold collar around the spot, drawing
in fluid from the near-surface regions to replace it.  The hot fluid
beneath the spot is positively buoyant; it is inhibited from rising
directly upwards by the magnetic field in the spot, and must therefore
first move axially outwards before it can rise around the spot.  It
collides with the upper descending cold collar, and the two are
deflected outwards away from the spot.  Some of the diverging fluid
then rises, and some of that then reconverges, producing a toroidal
eddy around the spot; the remainder of the ascending fluid is
deflected outwards, flowing away from the spot in the near-surface
layers.  That motion is quite difficult to perceive in Fig.~15,
which is but a single frame of a movie, for there are just two small
inclined arrows near each outer edge of the figure suggesting the
outward deflection.  But it is quite obvious when the movie is played.
However, that outward motion is not the Evershed flow.  It is too far
from the spot.  The structure of the visible spot is shown on the
representation of the upper horizontal boundary of the region being
depicted, and it is evident that immediately beneath the penumbra, and
somewhat beyond, the near-surface flow is axially inwards, towards the
spot.  This failure to miss the Evershed flow has spread considerable
doubt amongst solar physicists, particularly theorists and modellers,
on the reliability of the seismological inferences.  Perhaps that
doubt is justified. After all, Eddington said that one should never
trust an observation until it is confirmed by theory.  So I shall
address theoretical simulations in a moment.  But perhaps the doubt
was due as much to the reluctance of observers of only the superficial
layers of a star to accept more profound methods.  Ray Lyttleton once
said that if a modern observer were to meet a 
chimney sweep\footnote {It was commonplace in northern Europe up to
  half a century or so ago for houses to be heated by burning coal,
  often bituminous or the soft brown lignite coal which burns
  incompletely, encrusting the insides of chimneys with unwanted soot
  which subsequently might fall back into the room being heated or,
  more seriously, catch fire.  What escaped at the top of the chimney
  polluted the atmosphere, producing, under inclement conditions,
  dense unhealthy yellow-brown fog.  For safety, the soot had to be
  swept periodically from the insides of the chimneys, and a
  profession of chimney sweeps was established to perform that task.
  It was dirty work, and often a sweep's clothes and his exposed skin
  became covered with soot.  By contrast, a modern Danish chimney
  sweep prides himself of his cleanliness: he is well dressed, in
  tailcoat, top hat and white gloves.}  he would deduce that the sweep
were composed of pure carbon.

It is important to remain aware that, as I described when discussing
seismological inference of rotation, we cannot (readily) come to
reliable conclusions about conditions very near the solar surface from
the seismology of acoustic waves.  The top of Fig.~15 is about 2\,Mm
beneath the photosphere.  Therefore, if the situation presented by
that figure is correct, one must conclude that the Evershed flow is
shallow.

There is yet more seismological inference, which I have not yet
described.  In addition to acoustic waves there are surface gravity
waves, called f waves, whose physics is identical to that of the waves
on the surface of the ocean.  These waves do not propagate through the
interior of the Sun, but remain near the surface, their amplitudes
declining exponentially with depth at the same rate as they oscillate
horizontally (in other words, the e-folding depth is (2$\pi \rm
{)}^{-1}$ oscillation wavelengths).  They too are advected by flow.
Surface gravity waves confined essentially to a layer extending to
about 2\,Mm beneath the photosphere have been analysed by Laurent Gizon,
Duvall and Tim Larsen, who did indeed find outflow from the spot.  The
depth-averaged velocity is much less than that observed directly in
the photosphere, which is to be expected if the flow is a countercell
of the subsurface flow around the spot depicted in Fig.~15 whose
centre must lie less than 2\,Mm beneath the photosphere.  It seems that
these two complementary seismological analyses essentially complete
the basic picture.  I hasten to add, however, that that picture is not
accepted by a substantial number of theorists; Thomas and Weiss, for
example, consider such a shallow countercell to be unlikely.

It is evident from Fig.~15 that the subsurface inflow occurs in an
annulus that extends well beyond the penumbra.  So does the outflow
observed at the surface of the Sun, although the obvious penumbral
striations cease once the flow has passed the point at which it is
strongly influenced by magnetic field.  Therefore its superficial
appearance is different, and solar astronomers of late have given it a
different name: moat flow.  However, there appears to be no convincing
evidence that it is no more than simply the outer extent of the
Evershed flow.

Triggered by the doubt cast by solar physicists, helioseismologists
have reconsidered the approximations that were used in the
construction of Fig.~15: for example, the manner in which the
velocities observed at the ends of a ray-path segments (such as points
A and B in Fig.~14) are cross-correlated for inferring travel times,
the effect of ignoring the apparent time difference between the
reflection of an acoustic wave at its upper turning point and its
manifestation in the photosphere, the scattering by inhomogeneities
out of and into the ray path, diffraction, and the effect of
stratification on acoustic wave propagation.  All have some
quantitative impact on the inference, but at the moment it appears
unlikely that any is severe enough to make a qualitative change to the
picture.

There have been several attempts at direct numerical simulation of
sunspots.  Neal Hurlburt and Alastair Rucklidge have considered the
effect of a monolithic axisymmetric concentration of nearly vertical
magnetic field on convection in a layer of ideal gas.  In all cases
they found the fluid to converge on the field and sink in a cool
collar around the field, just as in Fig.~15.  They pointed out that
they had not modelled the solar atmosphere: they regarded the top of
their idealized model to be well below the solar photosphere, just as
are the current acoustic seismological inferences, and they too
embraced the idea that in the Sun there is a toroidal countercell
above the converging fluid which is manifest as the Evershed flow.
They also found an outer toroidal countercell surrounding the main
cell, which is diverging from the spot in its upper half, as is
(barely) seen in Fig.~15 (but is quite evident in the movie).
Hurlburt and Rucklidge suggested that that flow (without a countercell
above it) might be the moat flow.  The outflow evident at the upper
boundary of Fig.~15 (without a countercell above it) is so far from
the umbra that it could only be the outer extent of the moat.

The converging subsurface flow offers a natural explanation of how the
magnetic field is held together: it is continually advected inwards
against diffusion and its natural tendency to expand.  The superficial
layers that support the reverse Evershed flow have too little inertia
to offer significant opposition to that process.  In the deep layers,
below about 7\,Mm or so, the magnetic field has negligible influence on
the flow.  It surely seems most likely that the field is tangled by
the (three-dimensional) turbulent convection into thin flux tubes by a
process combining advection and diffusion akin to the pioneering
(two-dimensional) numerical studies carried out by Weiss in the 1960s.

\section{On the birth, death and lifespan of sunspots}

Sunspots tend to form in groups in regions in which there is a lot of
magnetic activity.  These regions are called, naturally enough, active
regions.  Active regions form, it is believed, from large magnetic
flux tubes that had been formed from field intensification possibly in
the tachocline beneath the convection zone, and have then risen
buoyantly to the surface.  The outcome is a pair of regions in which
the photosphere is crossed by magnetic field of opposite polarity,
moving away from each other and connected in an arch in the atmosphere
above, as in the cartoon depicted in Fig.~16.  This picture was
first adduced after studying the evolution of these regions from
observations of the photosphere and the overlying atmosphere; more
direct evidence for the rising of flux tubes before their appearance
at the surface has since been provided by seismology.  Active regions
can be up to 100\,Mm across.  They are temporary phenomena, with
lifetimes up to several months.  After an active region has
disappeared it is not uncommon for a new one to erupt in about the
same place, and on the longer timescale of several years there are
so-called active longitudes in the vicinity of which active regions
persistently form.  Understanding the long-term pattern of the coming
and going of these regions, which broadly indicate the locations of
the major sunspots in the butterfly diagram depicted in Fig.~6, is
the realm of solar dynamo theory.

\begin{figure}  
\begin{center}
  \includegraphics[width=0.5\textwidth]{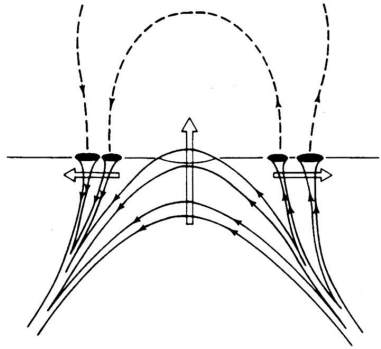}
  \caption[]{Cartoon of a rising flux tube, producing bipolar pairs of
    sunspots joined above by a magnetic arch (provided by Carolus
    Schrijver and Cornelis Zwaan).  The single arrows indicate the
    direction of the field, the double arrows the motion of the flux
    tubes.}
\end{center}
\end{figure}

The magnetic field that emerges in active regions is inhomogeneous,
initially being concentrated into flux tubes with cross-sectional
diameters of about 200km, containing field with intensity about 400G.
These tubes are quickly (on a timescale of less than an hour)
compressed by the convective flow into tubes 100km across with field
intensity about 1500G.  The tubes are advected by the supergranular
convection in such a manner as to cause them to meet, despite their
natural repulsive character, and coalesce into bigger tubes, called
pores, which sometimes, on a timescale of days, coalesce into yet
bigger tubes that then become fully fledged sunspots with penumbra.
The larger sunspots often form in recognizable pairs of opposite
polarity, joined by the magnetic arch in the atmosphere, although more
complicated groups, and individual sunspots, are not uncommon.  The
image of the solar atmosphere reproduced as Fig.~17 obtained from
the space mission TRACE tracing the magnetic field in an active region
near the solar limb is dominated by the field joining a large sunspot
pair.  But there is also more complicated magnetic structure which
undergoes reconnection, ejecting charged particles from the Sun and
creating what has been called space weather, which is a danger to
spacecraft and can upset the Earth's ionosphere, interfering with
radio communication and, in extreme circumstances, damaging power
lines.  Understanding the whole realm of these phenomena is now
sometimes referred to as heliophysics, although the word was
originally coined to encompass studies of only the (entire) Sun, from
the energy-generating core to the corona.

Sunspots of a pair are located very roughly east-west of each other,
which is consistent with them having risen from a field that has been
stretched into a toroidal coil in the tachocline, but inclined
somewhat such that the leading spot is closer to the equator.  The
inclination is a result of Coriolis torque (from a point of view in
the rotating Sun) as the field and its accompanying fluid moved
upwards and away from the axis of rotation -- that is simply the
tendency of the spot-pair to try to conserve its angular momentum,
thereby finding itself rotating more slowly than its surroundings.
Moreover, the relative polarities of the spots are opposite in the
northern and southern hemispheres, which is consistent with the idea
of tachocline winding of a basic large-scale internal dipole magnetic
field whose axis is aligned more-or-less with the axis of rotation.

\begin{figure}  
\begin{center}
  \includegraphics[width=0.8\textwidth]{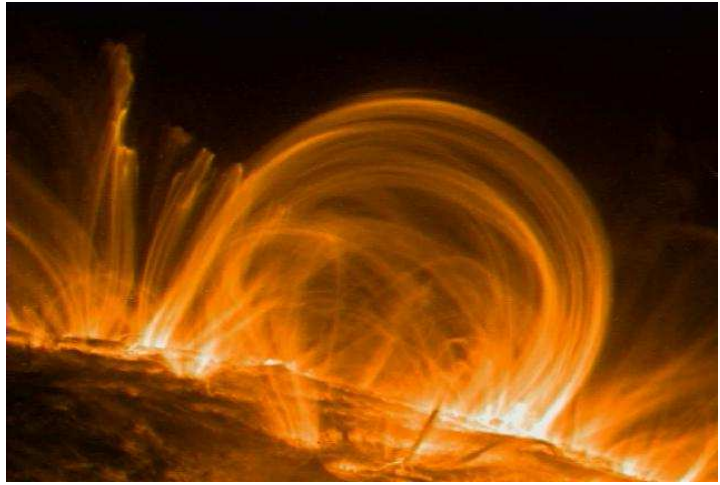}
  \caption[]{Image of an active region containing a large sunspot
    pair, taken by the spaceborne camera on TRACE.  The observation
    was made in the spectrum line H$\alpha$, which highlights the
    magnetic field (courtesy Alan Title).}
\end{center}
\end{figure}

As soon as a sunspot is created, it starts to decay.  The decay
appears to be consistent with the idea of lateral-surface abrasion by
the small-scale granular convection.  That is essentially a diffusive
process, and occurs much more slowly than sunspot formation -- large
sunspots are created in the course of days, but it then takes a month
or so for them to decline and die.  The timescale of diffusion scales
with the square of the linear dimension (it takes four times as long
to roast a turkey than it does to roast at the same temperature a
chicken of half the linear size: the roasting time of birds, or any
other food that scales homologously, is proportional to the two-thirds
power of the weight, contrary to the advice given in many cookery
books), and inversely with the magnitude of the diffusion coefficient.
If the diffusion coefficient of convective abrasion were constant,
the spot lifetime would be proportional to its area, and indeed there
is observational evidence corroborating that.  Not all spots are as
regular as those illustrated in Figs.~4 and 10, however; the scatter
in their properties is large, and the result of inferring any age--size
relation must be only approximate.  From some studies of the
observations it has been concluded that the effective diffusion
coefficient is proportional instead to the spot diameter.  When the
spot becomes small enough, it is essentially a pore.

According to this discussion, it is the convection that controls the
sunspot dynamics.  The same agent is responsible for both the birth
and the death of a spot.  How can that be?  Admittedly it is the
large-scale convection that appears to be responsible for a sunspot's
birth, and small-scale convection for its death.  But I have seen no
cogent explanation of why the large scales dominate in the early
stages of life and small scales in the decline -- so far as I
can see that has in most cases merely been implicitly assumed;
otherwise the matter appears to have been ignored.  Perhaps it is
simply a stochastic result moderated by the broad evolving conditions
in the active region.  There can be no sunspot decay in a spot-free
region; and a sunspot of any given size is more likely to be decaying
than be in a state of being created.  Perhaps that is simply because
the process of creation dominates the decay, but is only rarely
operational.  One is reminded of Boltzmann's $H$ theorem. Maybe it is
simply the very existence or not of a sunspot that biases future
evolution, just as statistical fluctuations in a stable thermodynamic
system are at any moment more likely to be decaying than growing,
causing entropy, on the whole, to increase.

\section{Solar-cycle irradiance variation}

I conclude by returning to the question of why it is that on a
solar-cycle timescale the solar irradiance at sunspot maximum, when
there is more direct darkening of the photosphere, is greater than it
is at sunspot minimum.  A partial answer to the apparent contradiction
has emerged from detailed studies by Peter Foukal, Judith Lean,
Judit Pap, Sami Solanki and their colleagues, who addressed
particularly the causes of shorter-term (daily-to-monthly) irradiance
variations evident especially at sunspot maximum.  They have found
that those fluctuations can be very well reproduced as a combination
of the reduced radiation from sunspots with enhanced radiation from
surrounding regions called faculae.  Faculae are structures in active
regions that are somewhat hotter than the normal atmosphere, being
hotter by about 100K in the photosphere and by substantially more than
that higher in the atmosphere.  They are closely associated with
sunspots, their total area following the solar cycle, roughly
preserving a facular-to-sunspot area ratio.  Being only slightly
hotter than the normal photosphere they are difficult to see near disc
centre, but they stand proud of the normal surface and are therefore
relatively more visible near the limb.  The radiation they emit
exceeds the sunspot deficit, which immediately explains why the
irradiance is greatest at sunspot maximum.  The extra energy that
heats them is presumably transported through the photospheric regions
directly by the magnetic field, rather than by convection and
radiative transfer, although some time ago I suggested, not without
(admittedly incomplete) theoretical justification, that a degree of
magnetic enhancement of convective transport under the photosphere of
the so-called quiet Sun (away from active regions) might also
contribute to enhanced irradiance, at least on solar-cycle timescales;
Gene Parker subsequently embraced this idea, at least for a while, but
on the ground that at the time no other plausible explanation could be
found.

\begin{figure}  
\begin{center}
  \includegraphics[width=0.32\textwidth]{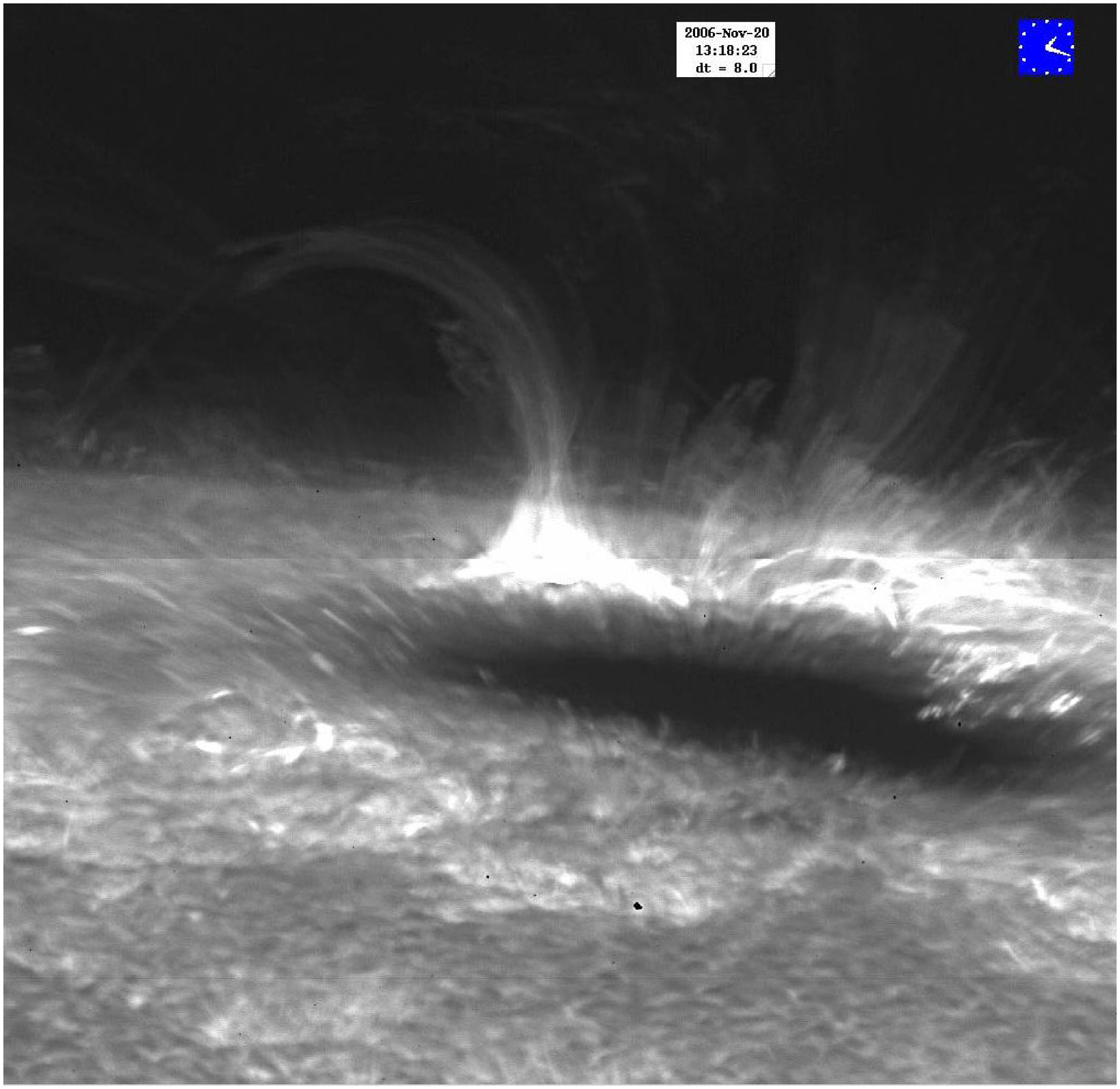}
  \includegraphics[width=0.32\textwidth]{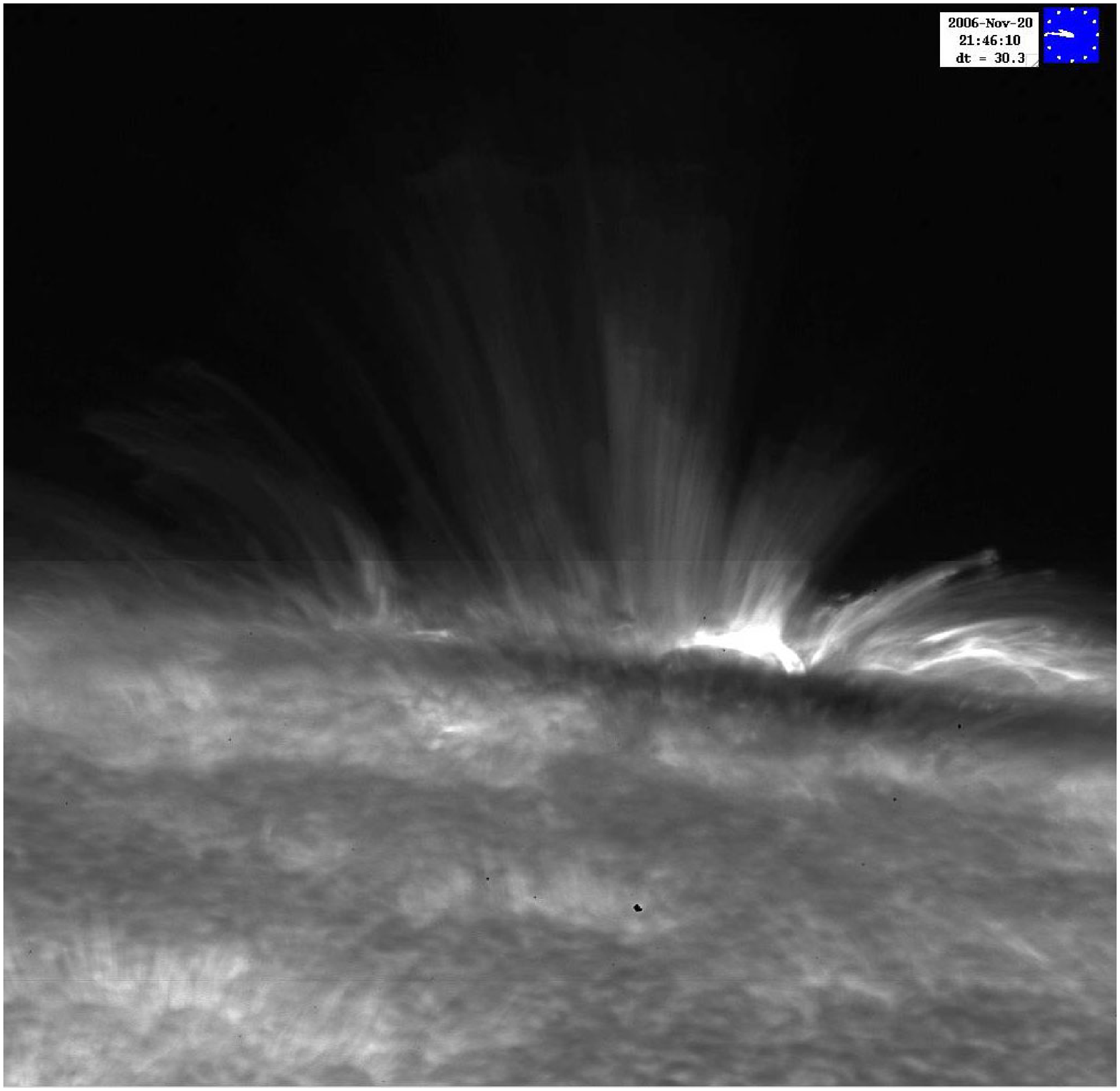}
  \includegraphics[width=0.32\textwidth]{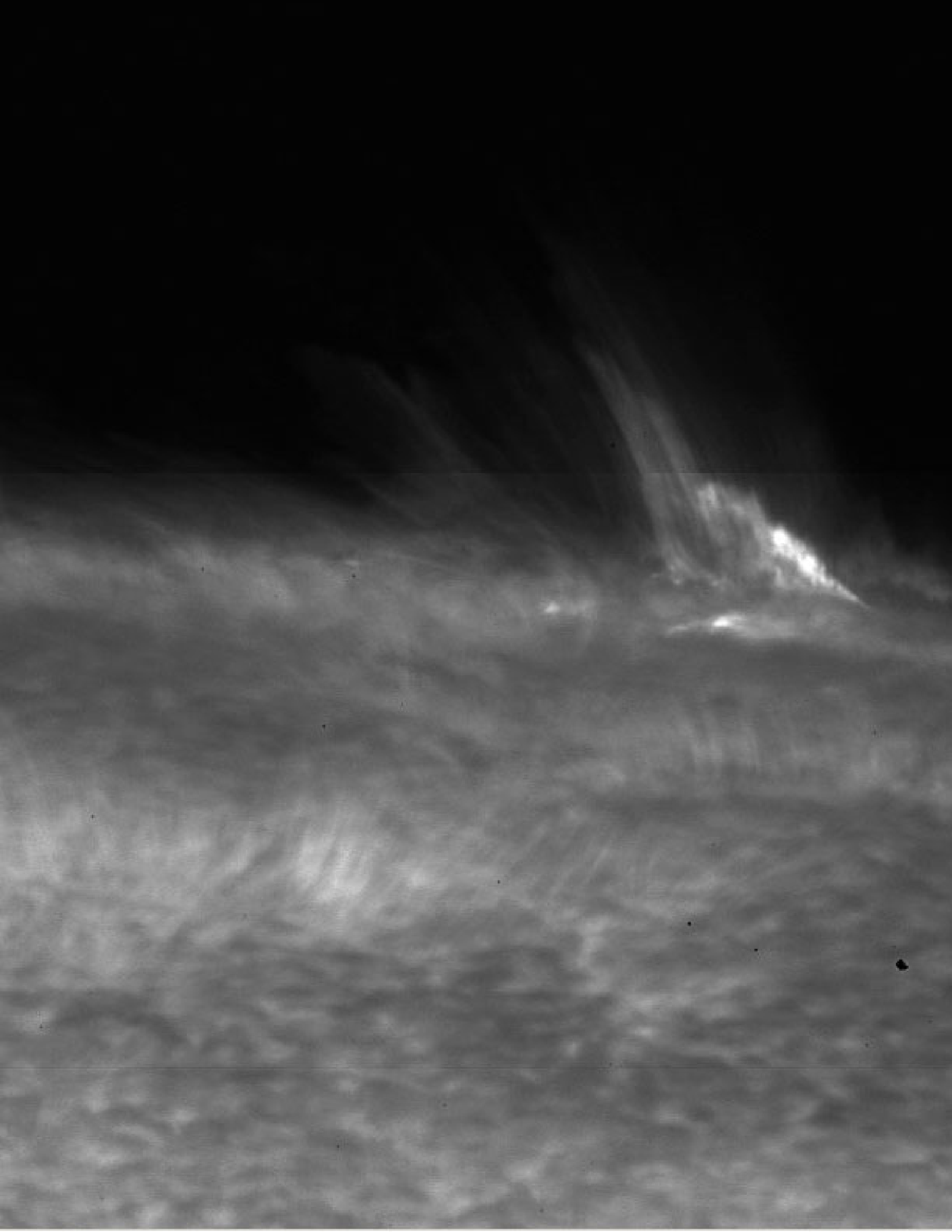}
  \caption[]{A sequence of three images of a sunspot approaching the
    solar limb on 20 and 21 November 2006, taken by the Hinode space
    mission.  Dates and times are indicated in the top right-hand
    corners.}
\end{center}
\end{figure}

An extreme example of what I am now talking about is beautifully
illustrated in a stunning movie taken recently from the Japanese
spacecraft Hinode (meaning Sunrise).  Fig.~18 is three images
selected from that movie.  In the first image the dark sunspot is
clearly visible; it is dominated by the umbra, which is evidently
lower than the surrounding photosphere.  Around it are bright faculae,
which extend up into the atmosphere.  This image is in contrast to
Figs.~4 and 10 depicting sunspots far from the limb, in which little
or no facula brightening is readily visible.  The faculae have become
relatively more prominent in the second image of Fig.~18, partly
because the depressed umbral region is substantially obscured by the
photosphere in the foreground, whereas the faculae remain fully
visible.  In the final image the spot, which is now quite close to the
limb, cannot be seen at all; but faculae are still apparent.  This
example not only reveals the facular brightening, but also
demonstrates that the angular distribution of the brightening is
different from that of both the normal photosphere and the sunspots.
Sunspot darkening is more evident when viewed from above; facular
brightening is more visible, relative to the photosphere, when viewed
from the side.  This causes the energy from the Sun to be radiated
anisotropically.  When viewed from the Earth the sunspots are, on
average, the most visible, because they lie in a band near the equator
(mainly between latitudes $\pm 30 \rm^o$ or so), which is close to the
plane of the ecliptic in which the Earth orbits.  If it were to be
viewed from the poles, however, the Sun would appear more luminous,
because the sunspots would be hardly visible yet a complete ring of
faculae would be seen shining in the vicinity of the limb.  Moreover,
after a little thought it is evident that when viewed from any
latitude away from the equator there is luminal enhancement, although
it is less than when viewed from the poles.  The amplitude of
solar-cycle luminosity variation, which is an integral over all
directions, is therefore greater than that of irradiance variation, by
about 30 per cent.  Of course it is only the irradiance variation that
concerns us directly on Earth, for it is that which controls the
overall energy budget of the Earth's atmosphere and has an influence
on climate.  But for those heliophysicists interested in the overall
energy budget of the Sun, it is the luminosity that counts.  In the
past it has been assumed, often without apparent caveat, that the
luminsoity variation is the same as the irradiance variation.  But it
is now evident that that is not so.

Finally, I should point out that it is only the temporal variation,
not the mean value, of the luminosity of the Sun that is significantly
affected by the sunspots and the faculae.  As in my discussion of
Ap-star spots, on a timescale exceeding the thermal relaxation time of
the convection zone -- about 10$^5$ years in the case of the Sun --
granted no other agent to produce temporal variation (I ignore the
main-sequence evolution of the Sun, which takes place on a
10$^{10}$-year timescale), the mean luminosity is determined by the
rate at which energy is supplied to the convection zone by the
radiative envelope, which is itself determined by the energy generated
by nuclear reactions in the core, and the outer layers of the Sun just
have to adjust to cope with the amount of energy flow required.  The
energy generation rate depends on the physical conditions in the core,
of course, which depends in turn on the weight of the envelope bearing
down on it, and on the value of the thermal conductivity of the poorly
conducting region beneath the convection zone.  Because the convection
zone has so little mass, any variation in its structure can have only
a very small influence on its weight, and therefore can cause only an
almost imperceptible change to the core, leaving the luminosity
essentially unchanged.

I have now travelled, rather hastily, from the surface of the Sun to
its central core, where the energy that powers the multitude of
magnetohydrodynamical processes in the directly observable surface
layers is produced.  Sunspots are but a single manifestation of these
processes, but one which has a long history, and which remains
incompletely understood.  There is still much research to be carried
out, even to acquire a firm understanding of flow discovered by John
Evershed one hundred years ago.

\begin{acknowledgement}
I thank Paula Younger for typing the manuscript, Guenter Houdek for
converting the figures into a format acceptable to \LaTeX  and for
assembling an early version of the powerpoint presentation used in the
delivery of the lecture, and Sacha Brun for powerpoint advice in its
editing.  I also thank Rob Rutten for converting this written version
of the lecture into the book format.
\end{acknowledgement}

\end{document}